# Effect of Requirements Analyst Experience on Elicitation Effectiveness: A Family of Empirical Studies

Alejandrina M. Aranda, Oscar Dieste, Jose I. Panach and Natalia Juristo

**Abstract**—*Context*. Nowadays there is a great deal of uncertainty surrounding the effects of experience on Requirements Engineering (RE). There is a widespread idea that experience improves analyst performance. However, there are empirical studies that demonstrate the exact opposite. *Aim*. Determine whether experience influences requirements analyst performance. *Method*. Quasi-experiments run with students and professionals. The experimental task was to elicit requirements using the open interview technique immediately followed by the consolidation of the elicited information in domains with which the analysts were and were not familiar. *Results*. In unfamiliar domains, interview, requirements, development, and professional experience does not influence analyst effectiveness. In familiar domains, effectiveness varies depending on the type of experience. Interview experience has a strong positive effect, whereas professional experience has a moderate negative effect. Requirements experience appears to have a moderately positive effect; however, the statistical power of the analysis is insufficient to be able to confirm this point. Development experience has no effect either way. *Conclusion*. Experience effects analyst effectiveness differently depending on the problem domain type (familiar, unfamiliar). Generally, experience does not account for all the observed variability, which means there are other influential factors.

---

## 1 INTRODUCTION

REQUIREMENTS elicitation is generally acknowledged as being one of the most important activities within the requirements engineering process [1], and it has a direct impact on software system quality [2]. It is an activity that requires intense communication between stakeholders (e.g., clients and analysts). Human interaction plays a critical role in this context [3].

There are a number of personal characteristics that may influence the effectiveness of any requirements-related task: experience [4], [5], [6], academic education [7], [8], cognitive capabilities [9], domain knowledge [1], [5], [10], [11], [12], [13], etc.

The idea that experience [14], [15], [16] improves requirements analyst performance is widespread in the software engineering (SE) community. However, empirical studies that experimentally research the effect of experience [4], [5], [6], [9], [17] have not been able to demonstrate that experience has a positive effect. Niknafs and Berry [5] even found that experience has a negative effect.

The aim of this article is to determine whether **requirements analyst experience influences requirements elicitation effectiveness**. To do this, we conducted a sequence of quasi-experiments analysing the effectiveness of requirements analysts depending on years of experience. The subjects that participated in the quasi-experiments were developers with different levels of experience recruited at the School of Computer Engineering, Universidad Politécnica de Madrid, and at an empirical fair as part of an international conference on software quality and requirements (REFSQ 2013). Subjects elicited requirements on two problem domains (one with which they were familiar and another with which they were not) using the open interview requirements elicitation technique.

Our results reveal that the effect of experience on analyst effectiveness varies depending on the problem domain type (familiar, unfamiliar) where the elicitation takes place. For the *familiar problem domain*, positive effects are observed for *interview experience* and, possibly, *requirements experience*. On the other hand, *professional experience* and *development experience* appear to have a negative effect. For the *unfamiliar problem domain*, none of the experience types has any effect. Our results also clearly indicate that, in both domains (familiar and unfamiliar), more experienced analysts are only slightly more effective than less experienced analysts.

The article is structured as follows. Section 2 reports the background and work related to this research. Section 3 describes the research methodology. Section 4 details the quasi-experiments conducted and reports the results, which are discussed in Section 5. Section 6 points out the main validity threats. Finally, Section 7 outlines the conclusions.

## 2 BACKGROUND

The study of experience goes back a long way. In the early days, its aim was to ascertain which factors make expert subjects perform better. Studies by [18], [19] revealed two main characteristics that experts had in common: thorough knowledge of, and long years of service in, their field of expertise. Experience is not related to any inborn talent, like intelligence, and it is very specialized, that is, experience is not transferable from one area to another [20].

---

*Alejandrina M. Aranda, Oscar Dieste, and Natalia Juristo are with the Universidad Politécnica de Madrid, 28660 Boadilla del Monte, Spain. Jose I. Panach is with the Universitat de València, 46010 València. E-mail: am.aranda@alumnos.upm.es, {odieste, natalia}@fi.upm.es, j.ignacio.panach@uv.es.*

TABLE 1 summarizes the major empirical studies addressing the influence experience related to engineering and requirements experience.

## 2.1 Studies of Software Engineering Experience

Experience has also been the subject of study in Software Engineering (SE) since its infancy. In the 1980s, the fields of interest were programming and low-level design [21]. Since then, experience has been studied in almost all SE areas: design [22], usability [23], testing [24], etc. Generally, the theory of experience proposed by [19] has been repeatedly confirmed. For example, experts recognize or remember more programming language sentences than novices [25], or employ a better reasoning or better problem-solving strategies than novices [26]. Expert subjects have also been proven to perform better in RE. Two[1] high quality studies, namely [14] and [27], confirm that expert analysts have specialized knowledge and use more complex reasoning than inexperienced analysts.

It is well-established that it takes about 10 years or 10,000 working hours to gain the required experience [28], [29], although this is a variable figure that depends on the area and type of instruction received [30]. In fact, shorter times are specified for SE. Campbell and Bello [31] claim that programmers need (at least) two years to become experts in Smalltalk. Sim et al. [32] consider that five years of experience is time enough for a software engineer to become an expert, which is the same figure as suggested by Atkins [33].

The passage of time is a necessary, albeit not sufficient, condition for gaining experience. Apparently novice subjects have quite often been found to outperform experts [28], [30], [34]. This has led to the time spent performing an activity (*experience*) being dissociated from proficiency at performing an activity (*expertise*). **An expert is anyone who outperforms** his or her colleagues, not a person with a long working history. Casual performance of an activity over a long period of time **is not sufficient** to achieve an expert level of skill; in turn, **expertise requires a prolonged period of intensive training** (e.g., the abovementioned 10 years) [35]. This explains why experience and expertise do not match [29], [36], [37].

To avoid confusion, we will use the terms *experience* or *experienced person*, and *expertise* or *expert*, according to their **restricted meanings** (as specified above) in the remaining of the paper.

The most common way of defining a SE professional's experience is **by years of experience** [38], as applies, for example, in job postings. It is not usual in SE for professionals to participate in lifelong training activities (that is, intensive training) once they have completed their higher education [32]. This suggests that experience and performance in SE should be at least partly correlated. Consequently, **we wonder how acceptable it is to use experience as a predictor of analyst effectiveness**. As analysts perform a variety of tasks, we focus on the requirements elicitation activity. This activity is not only important to assure the quality of the future software system, but also constitutes one of the biggest challenges for analysts [39]. It is reasonable to assume that experience, if it is clearly associated with better analyst performance or effectiveness during requirements elicitation, should be straightforward to observe empirically.

## 2.2 Studies about Requirements Engineering Experience

As far as we know, there are five empirical studies [4], [5], [6], [9], [17], all of which are of high quality, focusing on the relationship between analyst experience and effectiveness during requirements elicitation. Generally speaking, none of these studies have found experienced subjects to be more effective.

### 2.2.1 Pitts and Browne

Pitts and Browne [9] designed an experiment in the information systems field examining the use of cognitive stop rules in order to measure sufficiency, or the point at which the acquired requirements are sufficient for system development to continue.

A total of 54 professional analysts with at least two years of experience in systems development participated in the experiment. The average number of years of experience was 11. It is reasonable to assume that the subjects are mature enough to perform requirements elicitation effectively, i.e., according to some authors [31], they achieved the expert level.

Experimenters analysed the influence of experience (measured as number of years) in terms of number, breadth and depth of elicited requirements. As a result, as far as we are concerned, they reported that analyst experience does not influence requirements determination; that is, the number, breadth and depth of requirements does not depend on the number of years of analyst experience.

### 2.2.2 Marakas et al.

Marakas et al. [4] designed and ran a controlled experiment in the information systems field with the aim of evaluating the effectiveness of the semantic interview technique (a type of semi-structured interview) against the non-structured interviews.

A total of 20 inexperienced and experienced subjects participated in the experiment. Experience was measured as the number of years that the subject had worked on systems analysis and software development. Inexperienced subjects were final-year MSc in Software Engineering students, whereas experienced subjects were professional systems analysts and software developers.

The experimenters managed to identify differences in the effectiveness of the two interview types (specifically, subjects were more effective using the semantic interview). However, irrespective of the interview type used, experienced subjects were only marginally better (around 3%) than novice subjects, where the differences are nowhere near statistically significant.

---

[1] [14] may have used data taken from [27]. If this is the case, the two studies would count as only one.

TABLE 1
EMPIRICAL STUDIES OF REQUIREMENTS ENGINEERING EXPERIENCE

| STUDY | STUDY TYPE | SUBJECT | SAMPLE SIZE | TECHNIQUES | DEPENDENT VARIABLE | INDEPENDENT VARIABLE | | | | RESULTS | | |
|---|---|---|---|---|---|---|---|---|---|---|---|---|
| | | | | | | EXPERIENCE TYPE | EXPERIENCE RANGE | OPERATIONALIZATION OF LEVELS | EXPERIENCE EFFECT | SIG. | OBSERVATION | |
| Pitts & Browne [9] | Quasi-experiment | Systems analysts (professionals) | 54 | Unstructured interview | Effectiveness = Number of requirements | Systems development experience | >= 2 years | Years of experience | (+) Very low | r =.075 Sig. =.591 | Experience does not influence requirements elicitation. | |
| Agarwal and Tanniru [6] | Controlled experiment | Students + Professionals | 30 | Structured interviews Unstructured interviews | #rules(R) #criteria (C) | Novice subjects: postgraduate students | | Years of experience | | Mean Diff. | Experienced subjects tend to be marginally more effective (by 3.9%) than inexperienced subjects. However, this difference is not significant. | |
| | | | | | | Experienced subjects: Professional analysts | >= 1 year in expert systems projects >= 3 years in systems analysis | Expert Novice | (+) Very low (+) Very low | R = 0.9% C = 0.18% Non-significant | | |
| Marakas and Elam [4] | Controlled experiment | Students + Professionals | 20 | Semantic interviews Unstructured interviews | DFD scores | Development experience or systems analysis | | High Low | (+) Low (+) Low | r = .147 | Experience tends to have a positive effect. | |
| Niknafs and Berry [5] | Controlled experiment (E1) | Students | 12x3 | Brainstorming | #relevant (NR) ideas #feasible (NF) ideas | Professional experience | 0 to 4 years | Dichotomized: 0 years, 1-2 years and > 2 years | (−) | NR: p = .027* NF: p = .098** | Professional experience has a statistically significant negative influence on effectiveness (# ideas). | |
| | | | | | | Requirements experience | | | (−) | NR: p = .699 NF: p = .861 | Requirements experience has a statistically non-significant negative influence on effectiveness (# ideas). | |
| Niknafs and Berry [17] | Controlled experiment (E1+E2) | Students | 40x3 | Brainstorming | #relevant (NR) ideas #feasible (NF) ideas #innovative (NI) ideas #general (NG) ideas | Professional experience | 0 to 4 years | Dichotomized: None Low Medium High | (+) | NR: p = .583 NF: p = .059** NI: p = .559 NG: p = .461 | Professional experience partially influences the number of generated feasible ideas and does not have a significant influence on any of the dependent variables. | |
| | | | | | | Requirements experience | | | (−) | NR: p=.035* NF: p=.408 NI: p=.751 NG: p=.449 | Requirements experience tends to have a negative influence on effectiveness (relevant idea generation). | |

* Significant with α = 0.05; ** Significant with α = 0.10

### 2.2.3 Agarwal and Tanniru

Agarwal and Tanniru [6] conducted a controlled experiment in the expert systems field in order to compare the effectiveness of the structured and non-structured interview in terms of number of extracted rules, including other measures.

A total of 30 subjects with different levels of experience -novices and experienced- participated in the experiment. The novice subjects were postgraduate students with similar backgrounds and job experience, whereas the experienced subjects were either:

- Knowledge engineering professionals with job experience working on at least one expert system, or
- Analysts with at least three years of systems analysis experience.

The researchers reported that:

- Experienced subjects (who used unstructured interviews) performed slightly better (by about 9%) than inexperienced subjects (who used unstructured interviews). However, the differences were not statistically significant.
- Novice subjects (who used structured interviews) achieved better results than other novice and experienced subjects (who used unstructured interviews). The differences were statistically significant.

### 2.2.4 Niknafs and Berry

Niknafs and Berry [5] conducted a controlled experiment to empirically study the impact of domain knowledge and requirements experience on requirements elicitation effectiveness measured in terms of the number of generated ideas. Unlike previous research, the experimental subjects used brainstorming as the elicitation technique.

A total of 19 groups participated in the experiment. Each group was composed of three undergraduate subjects with different levels of development and requirements specification experience. Experience was measured in number of years. The experimenters argue that although it would make sense for there to be a positive relationship between experience and requirements elicitation effectiveness, the trends that they observed suggest quite the contrary, that is, subjects with one and two years of experience were slightly less effective than inexperienced subjects, whereas the effectiveness in groups with more than two years of experience dropped sharply.

Niknafs and Berry [17] reported two controlled experiments (E1 and E2) with computer science and software engineering students, E2 being an exact internal replication of E1 published in 2012 (Niknafs & Berry, 2012). The aim of E2 is to increase the sample size recruited in E1. A total of 40 groups participated, each with three members and differing levels of professional and requirements experience of up to four years. E2 was analysed together with E1. The results show that professional experience has a positive, albeit non-significant, effect on relevant, feasible, innovative and general idea generation. Requirements experience tends to have a positive, statistically significant, effect on idea generation. Otherwise, the effect is not significant.

## 2.3 Studies of Software Engineering Experience

Although there is a very widespread belief in SE that experience improves analyst effectiveness [14], [15], [16], [40], it is a fact that existing empirical studies [4], [5], [6], [9], [17] have failed to confirm this belief, sometimes even reporting results to the contrary. **The aim of this paper is to verify whether requirements experience influences analyst effectiveness during requirements elicitation**. To do this, we apply an empirical approach along the lines of the related work. Section 3 describes the study design.

## 3 METHODOLOGY

This section reports the experimental design, according to the SE experiment reporting guidelines proposed by [41].

### 3.1 Hypothesis

The main aim of the research is to experimentally analyse the influence of analyst experience on the effectiveness of the requirements elicitation process. To do this, we stated the following research hypothesis based on its respective null ($H_{0i}$) and alternative ($H_{1i}$) hypotheses:

$H_{0i}$: There is no relationship between experience and elicitation process effectiveness.
$H_{1i}$: There is a relationship between experience and elicitation process effectiveness.

being *i* one of the experience types defined below in Section 3.2. As our study is exploratory and the literature has reported different trends with respect to the effect of experience, we cannot anticipate the direction of the effects. Therefore, the alternative hypothesis is two-tailed.

### 3.2 Independent Variables

The independent variable refers to requirements analyst *Experience*, as shown in TABLE 2. However, it is immediately clear that the concept of "experience" is ambiguous (there is more than one type of experience: interview, requirements, etc.). It would not be very precise to consider just professional experience (subjects may have carried out many different activities throughout their career). On this ground, we decided to consider other types of experience apart from **professional experience**.

TABLE 2
INDEPENDENT VARIABLES

| INDEPENDENT VARIABLE | METRIC |
| --- | --- |
| • Interview experience<br>• Requirements experience<br>• Development experience<br>• Professional experience | Number of years of experience of each type |

Based on the related empirical literature, we specifically chose **requirements** ([6], [17]) and **development** ([9], [6]) **experience**. Although Marakas and Elam [4] do not clearly specify whether they study development or requirements experience, we construe analysis experience to be equivalent to requirements experience. We believe that it is important to also consider **interview experience**, because, as specified in Section 3.5, elicitation sessions were carried out by means of interviews, on which ground the subject's interview experience could influence their effectiveness.

## 3.3 Dependent Variables

The dependent variable is elicitation process *Effectiveness.* There is as yet no widely accepted metric for measuring requirements elicitation effectiveness. It is possible, however, to find several alternative measurements in the literature and existing empirical papers, where effectiveness is measured as the total number of identified rules [6]; according to the number of elicited requirements [9]; taking into account the elicited rules and clauses [42]; by dividing effectiveness into different categories [43] (total number of requirements, processes and information); and according to the number of ideas generated by the RE team [5].

Therefore, effectiveness has traditionally been operationalized as the number of items (irrespective of whether they are concepts, rules, processes, etc.) elicited by the analyst during elicitation. In this case, we applied a similar procedure to the abovementioned researchers and measured requirements elicitation process *Effectiveness* as the **percentage of all problem domain items identified by the experimental subjects**.

The problem domain is composed of different types of items. There is agreement in the literature on the essential items, such as objectives, processes and tasks [44], requirements, functions and states [45], concepts, actions or rules [46], etc.

Based on the above papers, we used three key types of items to measure effectiveness: concepts, processes and requirements. We did not take into account domain details (for example, process inputs and outputs, or conceptual model attributes and relations) because, as explained later, the experimental task is a short interview in which analysts would find it difficult to appreciate such details. We did not study more sophisticated aspects, like rules, objectives and stakeholders, either, as the selected domains are rather simple (with few stakeholders and no business rules). On the same grounds, non-functional requirements were excluded from the study. This has the advantage of making the experiment easier to instrument.

From the viewpoint of information systems, requirements are derived from the automation of pre-existing domain processes. Processes and requirements could be viewed as redundant for the purposes of measuring the amount of information gathered by subjects. However, other types of systems, e.g., control systems, do not share these peculiarities of information systems. In these contexts, a better option would be the characterization by [47], which regards requirements rather than specifications as operational declarations on a domain. From this standpoint, requirements are separate parts of the problem domain from processes and concepts, and should be addressed as such. This is the best option for this research, as the problem domains used have characteristics more closely resembling control systems than information systems.

The problem domain items considered in this research are: concepts, processes and requirements. The effectiveness measure is calculated according to the following formula:

$$Effectiveness = \frac{\frac{\#\ identified}{concepts} + \frac{\#\ identified}{processes} + \frac{\#\ identified}{requierements}}{total\ number\ of\ elements}$$

We opted not to calculate a weighted mean, as the fact that there is a greater percentage of a particular item in a domain does not mean that it is intrinsically more important. All the items play an important role in the construction of the future software system.

## 3.4 Subject Selection

We used convenience sampling to select the experimental subjects. The subjects that participated in the quasi-experiments were recruited from among either Requirement Engineering students enrolled in the Master in Software Engineering at the Universidad Politécnica de Madrid (UPM), or participants in the 2013 edition of the Alive Empirical Study at the International Working Conference on Requirements Engineering (REFSQ). Most experimental subjects (117 out of 124) have a computing or similar background.

## 3.5 Experimental Task

The experimental task was composed of three main phases: 1) elicitation session, 2) information reporting, 3) post-experimental questionnaire response. During the elicitation session, the subjects played the role of requirements analysts (interviewer), and three researchers acted as clients (interviewees).

The elicitation session was conducted by means of an open interview (that is, an open conversation as is common in the early stages of the requirements process) within a set time (30 or 60 minutes, depending on the case). We believe that the allocated times are sufficient to acquire a substantial amount of information about the problem domain.

At the end of the elicitation session, the experimental subjects submitted a written report about all the information that they acquired during the interview. There was no set reporting format, and they were given up to 90 minutes to complete the report. Although templates are the most common requirements reporting format, we observed that the experimental subjects tended to report a list of items, often divided into sections (e.g., functional requirements, non-functional requirements, etc.) and even managing to use conceptual models. Therefore, we decided to allow subjects to use their preferred reporting format instead of requiring template use.

At the end of the experiment and after submitting their final report, the subjects completed the post-experimental questionnaire, which took fewer than five minutes.

## 3.6 Assigment of Subjects to Treatments

We used a quasi-experimental design. Quasi-experiments are carried out when the subjects cannot be assigned at random to an experimental condition or, alternatively, a treatment cannot be assigned to a group. This is applicable here, since the experience is a built-in characteristic of the experimental subjects that cannot be randomized or blocked.

Therefore, subjects are not, strictly speaking, assigned to treatments in this study, because they all perform the same treatment, that is, all the subjects participating in the quasi-experiment perform requirements elicitation on the same problem and with the same interviewee.

### 3.7 Experimental Objects

In this research, we used two problem domains, as shown in TABLE 3. In the first case, we used a domain with which the analysts were unfamiliar, that is, the selected problem is so uncommon that most experimental subjects are unlikely to have any exposure to the issue. This rules out an uncontrolled *Experience x Knowledge* interaction[2]. The unfamiliar problem domain (UP1) is related to a battery recycling plant, where a series of domain-specific machines perform several peculiar processes practically impossible to infer unless one has first-hand knowledge of such domain. The problem is based on a simplified real system to assure that the subjects can address the problem in the limited time available for the experimental sessions.

The familiar problem domain (FP1) is an instant messaging system by means of which the user will be able to perform operations like flat text messaging, contact management, etc. In this case an *Experience x Knowledge* interaction is possible, but such effect can be estimated by comparison with UP1.

Both problem domains were described in full according to three types of items by which they are defined: requirements, concepts and processes as shown in Appendix B. These descriptions are useful for training interviewees to provide the right answers to interviewers. They can be also used as a yardstick for measuring the effectiveness of the experimental subjects. TABLE 4 shows the total number of items defining the size of the problem domain. Note that we tried to assure that the total number of problem items was similar. This was intended to make sure that the *Problem* was not another variable possibly influencing the results, as, otherwise, we would have an unwanted *Knowledge x Size* interaction.

TABLE 3
PROBLEM DOMAINS USED IN THE EXPERIMENT

| PROBLEM | BRIEF DESCRIPTION | TYPE |
|---|---|---|
| UP1 | Battery recycling machine control system. | Unfamiliar |
| FP1 | Text messaging system. | Familiar |

TABLE 4
TOTAL NUMBER OF ITEMS FOR EACH PROBLEM DOMAIN

| PROBLEM | ITEMS DEFINNING THE PRBLEM (#) | | | |
|---|---|---|---|---|
| | REQUIREMENTS | CONCEPTS | PROCESS | TOTAL |
| UP1 | 15 | 24 | 12 | 51 |
| FP1 | 28 | 10 | 16 | 54 |

### 3.8 Measurement Procedure

The independent variables were gathered by means of a post-experimental questionnaire. It was implemented within the *Moodle* and *Google Forms* platforms. The aim of the questions was to gather information related to the experimental subject: years of experience, knowledge on problem domains and training, etc. The questionnaire is available in Appendix A.

The dependent variable, elicitation process effectiveness, was measured according to the reports submitted by the experimental subjects at the end of the reporting process. The items defining the problem domain (requirements, concepts and processes) were used as a checklist to measure effectiveness. Multiple repetitions of the same item in reports submitted by subjects were not taken into account to calculate effectiveness. Each quasi-experiment was measured by a single researcher. This could lead to some bias in the results and is considered as a possible validity threat in Section 6.

### 3.9 Analysis Strategy

Typical inference tests (t, ANOVA, etc.) cannot be applied in the analysis of the quasi-experiments, as such tests cannot use scalar independent variables. One alternative option is an ANCOVA using the domain type as a between factor and experience as a covariable. However, Aranda et al. [48] already evaluated the effect of knowledge, which is not the focus of this paper. Additionally, as mentioned later in Section 4.4, groups are unbalanced, and one of the data sets is possibly heteroscedastic, which advises against the use of ANCOVA. We could use mixed models, but we would have to assume a particular covariance matrix structure, which is more complicated to interpret.

We have decided to use multiple linear regression (MLR). This statistical technique is appropriate because: a) it can account for scalar independent variables, and b) it is capable of jointly calculating the effects of several independent variables. For MLR to be reliable, we have to test for four conditions: collinearity, sampling size, normality and homoscedasticity.

- **Collinearity.** For the analysis to be reliable, it is necessary to assure that the model predictor variables are not collinear. Collinearity occurs in MLR when one or more independent variables are linearly correlated with other model variables. To check for collinearity between variables, we used the variance inflation factor (VIF), tolerance (T) and condition index (CI):
  - One recommendation often used by researchers [49] is to consider a large VIF, that is, $VIF > 10$, yielded if $R^2 > 0.9$ and $T < 0.1$, as evidence of collinearity. A second more rigorous option is to reduce the VIF limits to $VIF > 5$ with $R^2 > 0.8$ and $T < 0.2$ [50].
  - *Condition index (CI):* Belsley [51] suggests three degrees of collinearity: slight ($CI < 10$), moderate ($10 < CI < 30$) and severe ($CI \geq 30$). When a model has a severe condition index, one or more variables have shared variance with the other varia-bles. Usually, a variable with a high variance proportion (greater than 0.5) is considered to be involved in collinearity.
- **Normality.** The distribution of residuals must be normal with a mean of zero and random but constant variance. We used the Kolmogorov-Smirnov test with the Lillierfors correction and the Shapiro Wilks test to check for normality of residuals.
- **Homoscedasticity.** We checked for homogeneity of variance using scatter plots of the standardized predicted values against model residuals.

---

[2] Notice that the impact of problem domain knowledge is not one of the research goals of this paper; we have already reported about this issue in [48]. Our intention behind the separation of the *Experience* and *Knowledge* variables is to avoid the threat of previous problem domain knowledge having an influence on analyst effectiveness that is confounded with the experience effect.

We interpret the effect of experience using the non-standardized coefficient (B). B indicates the mean change for the dependent variable against each unit of change of the independent variable, that is, the resulting Bs represent increases (or decreases) in effectiveness by year of experience. Bs are reported using the same unit as the response variable *Effectiveness*, that is, percentages. For example, five years of experience for a B = 1.543 is equivalent to an effectiveness increase of 1.543 × 5 = 7.7%. Unlike effect sizes, we have no benchmarks for determining whether a particular B is equivalent to a large, medium or small effect. On this ground, we will set a criterion for interpreting the Bs as follows in TABLE 5.

### 3.10 Sample Size Estimation

To be able to rigorously apply a MLR, there should be a minimum sample size. Otherwise, the estimation of the effect of the independent variables will be less precise, the likelihood of detecting significant effects will be lower, and there may even be an overfitting phenomenon. Overfitting occurs when the model provides too exact a correspondence with the data set because there are too many independent variables for the number of cases. According to Miles and Shevlin [52] (cited in Field et al. [53]), 90 subjects are enough to test a MLR with four independent variables, assuming medium effect sizes. If the effects were large, 40 subjects would suffice.

## 4 FAMILY OF QUASI-EXPERIMENTS

### 4.1 Quasi-experiments

We conducted 12 quasi-experiments: eight on the unfamiliar problem domain (UP1) and four on the familiar problem domain (FP1).

TABLE 6 shows the quasi-experiments conducted using the UP1 problem domain. For each quasi-experiment, it shows the type of replication used, the execution site and the number of subjects participating in each replication. Q-2007 was the baseline experiment, which we replicated seven times: six replications at UPM and one at REFSQ. We collected data about 88 experimental subjects.

To study the effect of experience in the familiar domain, we executed four quasi-experiments with a total of 36 experimental subjects, shown in TABLE 7.

Comparing TABLE 6 and TABLE 7 it can be easily noticed that the research started in the unfamiliar domain (quasi-experiment #1 dates back to 2007), and only much later (2012) progressed to the familiar domain.

TABLE 5
TOTAL NUMBER OF ITEMS FOR EACH PROBLEM DOMAIN

| EFFECT | LARGE | MODERATE | LOW | ZERO |
|---|---|---|---|---|
| Coefficient (B) | ± (>= 3) | ± [2 - 3] | ± [1 - 2] | ± [0 - 1] |

TABLE 6
FAMILY OF EMPIRICAL STUDIES ABOUT REQUIREMENTS ELICITATION – UP1

| # | QUASI-EXPERIMENT | REPLICATION TYPE | SITE | SUBJECTS |
|---|---|---|---|---|
| 1 | Q-2007 | Baseline experiment | UPM | 7 |
| 2 | Q-2009 | Internal replication | UPM | 8 |
| 3 | Q-2011 | Internal replication | UPM | 16 |
| 4 | Q-2012-REFSQ | External replication of Q-2011 | REFSQ | 21 |
| 5 | Q-2012-UP1 | Internal replication | UPM | 14 |
| 6 | Q-2013-UP1 | Internal replication | UPM | 8 |
| 7 | Q-2014-UP1 | Internal replication | UPM | 9 |
| 8 | Q-2015-UP1 | Internal replication | UPM | 5 |
| TOTAL | | 8 empirical studies | UPM / REFSQ | 88 |

TABLE 7
FAMILY OF EMPIRICAL STUDIES ABOUT REQUIREMENTS ELICITATION – FP1

| # | QUASI-EXPERIMENT | REPLICATION TYPE | SITE | SUBJECTS |
|---|---|---|---|---|
| 9 | Q-2012-FP1 | Internal replication | UPM | 14 |
| 10 | Q-2013-FP1 | Internal replication | UPM | 7 |
| 11 | Q-2014-FP1 | Internal replication | UPM | 7 |
| 12 | Q-2015-FP1 | Internal replication | UPM | 8 |
| TOTAL | | 4 empirical studies | UPM / REFSQ | 36 |

Our initial intention was to assess the experience effect independently of the problem domain knowledge effect (see Section 3 for details).
However, after conducting quasi-experiments #1-4, we observed that experience did not seem to make any influence on analyst effectiveness (see Section 4.4). Accordingly, we launched a series of quasi-experiments in a familiar domain, aiming to find out if the interaction *Experience x Knowledge* could explain the widespread belief in the positive effect of experience. The research finished when we approached the required sample sizes (see Section 3.10) for both domains, and the effects became apparent.

As shown in TABLE 8, quasi-experiments differed with respect to resource availability and contextual issues throughout the research. These differences have to be taken into account, as they can have a moderator effect on analyst effectiveness, that is, increase or decrease their effectiveness. For example, as TABLE 8 shows, the subjects in Q-2007 conducted the elicitation using the individual open interview as an elicitation technique, whereas the group interview (modelled on a requirements workshop) was used in Q-2011. It is evident that the interview type (individual, group) could influence the effectiveness achieved by analysts.

TABLE 8
QUASI-EXPERIMENT CHARACTERISTICS

| EXPERIMENT | INTERVIEW TYPE | LANGUAGE | INTERVIEWEE | ELICITATION TIME | EXECUTION | WARMING-UP |
|---|---|---|---|---|---|---|
| Q-2007 | Individual | Spanish | OD | 30 min | At the end of the course | 16 weeks |
| Q-2009 | Individual | English | AG | 30 min | At the end of the course | 16 weeks |
| Q-2011 | Group | English | OD | 60 min | At the end of the course | 16 weeks |
| Q-2012 REFSQ | Group | English | OD | 60 min | - | No warming up |
| Q-2012 | Individual | English Spanish | OD / JWC | 30 min | At the start of the course | No warming up |
| Q-2013 | Individual | English Spanish | OD / JWC | 30 min | At the start of the course | Warming up 1 week |
| Q-2014 | Individual | English | OD | 30 min | At the start of the course | Warming up 6 weeks |
| Q-2015 | Individual | English Spanish | OD / JWC | 30 min | At the start of the course | Warming up 2 weeks |

The contextual variables that could act as moderator variables between the analyst effectiveness and the years of experience are:

- **Interview Type:** individual open interview (1:1 interview) or group interview (1: N interview, where several analysts interview a client simultaneously).
- **Interviewee:** person who acts as the client during the requirements elicitation process. Three researchers played the role of client: OD (Oscar Dieste), AG (Anna Grimán) and JWC (John W. Castro).
- **Language:** some quasi-experiments are blocked by language (that is, one group held the interview in Spanish and another in English) and by interviewee. By blocking the subjects by language/interviewee, we increased the quality of the conversation (for example, a non-Spanish speaker could not communicate well enough in Spanish with a Spanish-speaking interviewee).
- **Elicitation Time:** interview duration. The interviews lasted at most 30 minutes and 60 minutes for individual and group interviews, respectively.
- **Execution:** whether subjects participated in the quasi-experiment before or after the Requirements Engineering course.
- **Warming-Up:** short training course in requirements-related activities. Warming-up refers to the duration of the training, ranging, in this case, from 0 to 6 weeks, whereas the above *Execution* moderator variable is essentially a binary variable (with *before* and *after* values).

### 4.2 Data Collection

#### 4.2.1 Demographic Data

TABLE 9 summarizes the key demographic data of the sample. We categorized the experience of the subjects at three levels: novices (0-1 year), intermediate (2-4 years) and experts (≥ 5 years) [31], [32], [33]. We found that the distribution of experience for UP1 is reasonably balanced among novices, intermediate levels and experts. This does not apply for FP1, and unbalance may have an impact on the reliability of the results for the familiar domain, as specified in the validity threats. For a more thorough examination of the distribution of experience, Appendix C provides bar charts by domain and experience type.

TABLE 9
CHARACTERIZATION OF SUBJECTS DEPENDING ON EXPERIENCE

| FAMILY OF QUASI-EXPERIMENTS | | | |
|---|---|---|---|
| CHARACTERISTIC | LEVEL | #UP1 SUBJECTS | #FP1 SUBJECTS |
| Academic Education | NCS | 7 | 0 |
| | CS | 80 | 34 |
| Interview Experience | Novices (0-1 year) | 53 | 24 |
| | Intermediate Level (2-4 years) | 17 | 6 |
| | Experts (>= 5 years) | 13 | 1 |
| Requirements Experience | Novices (0-1 year) | 44 | 21 |
| | Intermediate Level (2-4 years) | 22 | 9 |
| | Experts (>= 5 years) | 17 | 2 |
| Development Experience | Novices (0-1 year) | 13 | 8 |
| | Intermediate Level (2-4 years) | 30 | 15 |
| | Experts (>= 5 years) | 24 | 6 |
| Professional Experience | Novices (0-1 year) | 13 | 9 |
| | Intermediate Level (2-4 years) | 20 | 9 |
| | Experts (>= 5 years) | 38 | 9 |

NCS: Non-computerscience, CS: Computer science

#### 4.2.2 Descriptive Statistics

TABLE 10 shows the key descriptive statistics. There are no clear observable trends, as averages follow a sawtooth pattern. Exceptionally, there appears to be an increase in the effectiveness of the subjects depending on interview experience and a decrease depending on professional experience in both UP1 and FP1. The experimental data are available in http://grise.upm.es/sites/extras/15.

#### 4.2.3 Adaptation of Experimental Data

The possible effects of the contextual variables should be accounted for in order to conduct a more accurate joint analysis of the data from the different quasi-experiments. For example, imagine that we have two subjects, one non-computer scientist (A) and one computer scientist (B). The only difference between A and B is that B is familiar with the application domain. Likewise, this knowledge acts as a moderator variable and increases his or her performance by 10 points (see Fig. 1a). If we were to analyse the relationship between analyst performance and academic education, omitting the knowledge effect, the relationship would be positive (r>0). This result is incorrect, precisely, because the moderator variable (knowledge) is what really makes subject B more effective. The effect of this variable is confounded with the effect of the real independent variable (qualification), leading to the mistake in the estimation of the correlation coefficient.

TABLE 10
DESCRIPTIVE STATISTICS

| CHARACTE-RISTIC | LEVEL | UP1 | | | FP1 | | |
|---|---|---|---|---|---|---|---|
| | | #SUB | MEAN | STD. DEV | #SUB | MEAN | STD. DEV |
| Academic Education | NCS | 7 | 35.29 | 14.32 | 0 | | |
| | CS | 80 | 42.60 | 16.44 | 34 | 27.09 | 16.04 |
| Interview Experience | 0-1 year | 53 | 44.25 | 16.92 | 24 | 33.02 | 18.60 |
| | 2-4 yrs | 17 | 36.22 | 18.47 | 6 | 39.20 | 19.01 |
| | >= 5 yrs | 13 | 42.68 | 9.38 | 1 | 55.56 | . |
| Requirements Experience | 0-1 year | 44 | 44.30 | 15.40 | 21 | 36.11 | 21.75 |
| | 2-4 yrs | 22 | 38.68 | 19.57 | 9 | 37.65 | 20.47 |
| | >= 5 yrs | 17 | 41.64 | 14.33 | 2 | 30.56 | 18.32 |
| Development Experience | 0-1 year | 13 | 40.27 | 15.55 | 8 | 36.11 | 21.75 |
| | 2-4 yrs | 30 | 44.58 | 18.11 | 15 | 37.65 | 20.47 |
| | >= 5 yrs | 24 | 42.57 | 14.60 | 6 | 30.56 | 18.32 |
| Professional Experience | 0-1 year | 13 | 46.15 | 11.92 | 9 | 36.42 | 19.49 |
| | 2-4 yrs | 20 | 47.65 | 17.71 | 9 | 40.95 | 24.58 |
| | >= 5 yrs | 38 | 40.35 | 16.24 | 9 | 34.98 | 15.20 |

NCS: Non-computerscience, CS: Computer science, SUB: Subjects

The best way to account for the effects of moderator variables would be to add them as interaction terms to the MLR model discussed in section 4.3. However, this strategy could be troublesome for two main reasons:
1) Each new variable entered in a MLR model increases the sample size required to achieve reasonable statistical power. We have six contextual variables presented in Section 4.1: Interview type, intervieweelanguage, elicitation time, execution and warming up, although two of them have a 1:1 relationship (Interviewee - Language and Interview type - Elicitation time). Added to the four independent variables (the four types of experience: interviews, requirements, development, professional), it makes a model containing 8 variables. The required sample size to detect medium effect sizes would be around 125 subjects, according to Miles and Shevlin [52]. Our data set is not large enough.
2) From the viewpoint of the regression model, moderator variables are not different from independent variables. On this ground, orthogonality and a reasonable balance between combinations should exist in order the MLR be accurate. As the contextual variables in this research emerge on the grounds of contextual restrictions rather than design considerations, orthogonality and balance do not hold.

To solve this problem, we estimated, based on the available studies, different values for the moderator variables. We then averaged these values depending on the sample size of each study (as in meta-analysis, where sample size acts as a proxy of study reliability). We are reasonably sure that the applied adjustment procedure is reliable. Sometimes, the effect of a moderator variable can be estimated by means of a statistical model.

When this happens, the results of both procedures (statistical model, weighted averages) are very similar. For example, the interviewee effect was obtained using a mixed linear model in one particular study [48]; this value was 26%. The same interviewee effect after applying the weighted averages ranged from 18 to 23%. Therefore, the values are quite similar to each other.

In this manner, the influence of the moderator variables can be eliminated by subtracting the estimated effect of each variable on the affected subjects. Similarly, the performance of subjects A and B in the above example (see Fig. 1.b) is similar if we subtract the effect of the moderator variable for subject B (i.e., 10 points). This is what we would expect if we assume that there is no difference between the two. The data adjustment values for each moderator variable are specified in TABLE 11 The data adjustment procedure, as well as the details of the calculations are specified in Appendix D. The data adjustment does not change the patterns show in TABLE 10.

### 4.2.4 Data Set Reduction
For MLR, data must be available on all the experience types (interview, requirements, development, professionnal) for all subjects. As TABLE 9 shows, a number of subjects did not provide information on at least one of the types of experience. To apply MLR, these subjects must be removed. This means that:
- **Unfamiliar domain:** We recruited 88 experimental subjects, of which only 69 could be used.
- **Familiar domain:** Of 36 participant subjects, only 29 could be used.

Data loss can be reduced using fewer variables in the MLR. For example, if we take development experience out of the analysis, the usable data are greater for UP1 (76 subjects) and constant (29 subjects) for FP1. We did not adopt this approach as the models with three or four variables essentially yield the same results, as shown in Appendix E.

## 4.3 Testing of the Necessary Conditions for Applying MLR
As we studied the effect of experience on two different problem domains, the conditions of MLR were tested depending on each domain.

### 4.3.1 Unfamiliar Problem Domain
The MLR for studying the effect of different types of experience is composed of the four independent variables under study, namely, interview experience, requirements experience, development experience and professional experience.

$$Effectiveness = \beta_0 + \beta_1\ InterExp + \beta_2\ ReqExp + \beta_3\ DevExp + \beta_4\ ProfExp + \varepsilon.$$

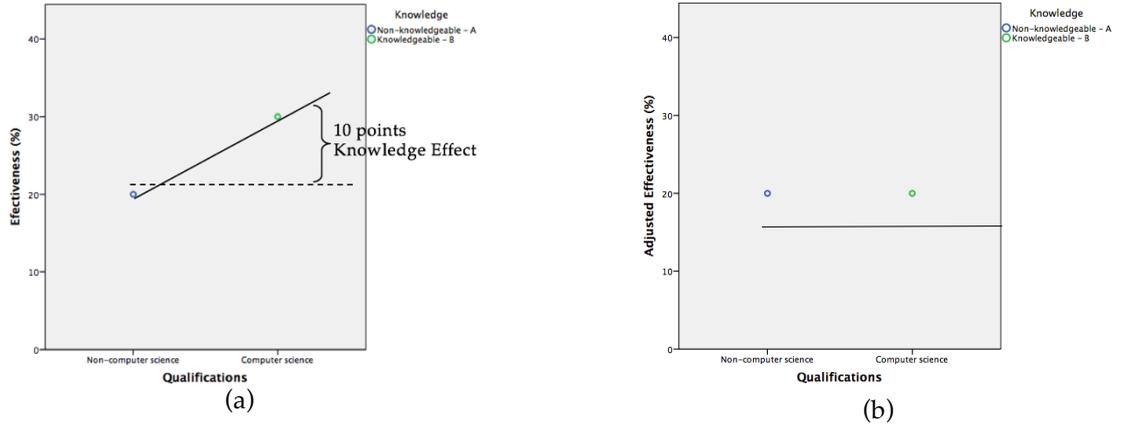

Fig. 1. Fictitious example of the effect of a moderator variable

TABLE 11
PERCENTAGE ADJUSTMENT (EFFECT) FOR EACH POTENTIAL MODERATOR VARIABLE

| MODERATOR VARIABLE | LEVELS | % ADJUSTMENT FOR UP1 | % ADJUSTMENT FOR FP1 |
|---|---|---|---|
| Interview type | Individual | 11% | NA |
| | Group | | |
| Execution | Before | 12% | NA |
| | After | | |
| Interviewee | OD | 23% | NA |
| | AG | | |
| | OD | 18% | 23% |
| | JWC | | |
| Warming Up | 0 week | 0% | 0% |
| | 1 week | -1% | -6% |
| | 2 week | 7% | 0% |
| | 6 week | 10% | 5% |

The proposed model meets all the required assumptions (see Section 3.9) for applicability:

- **Non-Collinearity:** Collinearity statistics are within the specified ranges (VIF < 10 and CI < 10), as shown in Section 4.4.1, TABLE 12 and Appendix F.1, Table 10, respectively.
- **Normality:** The distribution of the model residuals is normal (p-value = 0.200 > 0.05 for the Kolmogorov-Smirnov test and p-value = 0.243 > 0.05 for the Shapiro-Wilk test). Data normality is confirmed by means of skewness (-0.538) and kurtosis (0.233) statistics, as they are within the usual ranges ±1.
- **Homoscedasticity:** The observed variance is quite uniform across the range of typified residuals, as shown in Fig. 2. No bottleneck patterns are observed. The sample size is lower than the desired value. The model has four independent variables, on which ground at least 90 subjects are required to detect a medium-sized effect. In this case, the sample size (n=69) is not large enough; however, it comes very close and it is sufficient to detect medium-to-large effect sizes (see Section 3.10). The analysis results (see Section 4.4.1) point out in the direction that the low sample size is not affecting negatively the conclusions of this research. In any case, the low sample size is listed as a potential threat to validity.

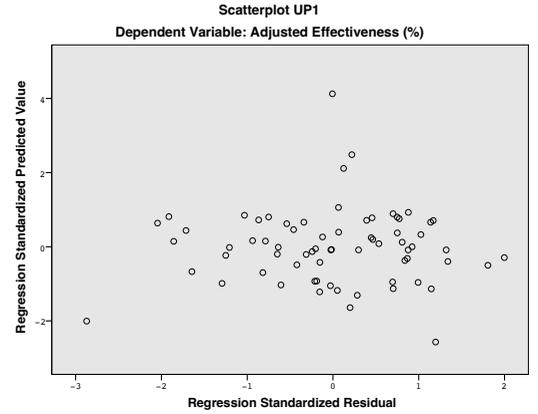

Fig. 2. Scatter plot of the residuals of the UP1 model

### 4.3.2 Familiar Problem Domain

The MLR for the familiar domain is the same as above. The data compliance with the applicability conditions is reasonable:

- **Non-Collinearity:** The collinearity statistics are within the established ranges (VIF < 10 and CI < 10), as specified in Section 4.4.2, TABLE 13 and Appendix F.2, Table 11, respectively.
- **Normality.** The model residuals have a normal distribution (p-value = .200 > 0.05 for Kolmogorov-Smirnov test and p-value = .453 > 0.05 for Shapiro-Wilks test). Data normality is confirmed by means of skewness statistics (.623) and kurtosis (.580), as they are within the usual ranges ±1.
- **Homoscedasticity.** As not many data are available, the patterns shown in Fig. 3 are not well-defined. On this ground, we cannot state or reject the homogeneity of variance. The missing information is discussed in the validity threats.

As in the unfamiliar domain, the sample size is not enough. The model has four independent variables, on which ground at least 40 cases are required to detect large effect sizes. In this case, we have n=29 subjects. This value is nowhere near the requirements. The MLR reported here will only be able to detect large effects with a power of 80% and $\alpha = 0.10$ (calculations made using G*Power).

TABLE 12
EFFECT OF EXPERIENCE FOR THE UP1 – MRL

| Model | Unstandardized Coefficients | | Standardized Coefficients | T | Sig. | Collinearity Statistics | |
|---|---|---|---|---|---|---|---|
| | B | Std. Error | Beta | | | Tolerance | VIF |
| 1 (Constant) | 28.730 | 2.543 | | 11.297 | .000 | | |
| Interview Experience (years) | .532 | .785 | .149 | .678 | .500 | .318 | 3.145 |
| Requirements Experience (years) | .271 | .717 | .102 | .378 | .707 | .209 | 4.787 |
| Development Experience (years) | .231 | .524 | .072 | .441 | .661 | .574 | 1.742 |
| Professional Experience (years) | -.540 | .606 | -.235 | -.891 | .376 | .221 | 4.535 |

Dependent Variable: Adjusted Effectiveness (%)  $R^2$ = .020; Real N = 69; Required N = 82

TABLE 13
EFFECT OF EXPERIENCE FOR THE FP1 – MRL

| Model | Unstandardized Coefficients | | Standardized Coefficients | T | Sig. | Collinearity Statistics | |
|---|---|---|---|---|---|---|---|
| | B | Std. Error | Beta | | | Tolerance | VIF |
| 1 (Constant) | 29.247 | 5.823 | | 5.023 | .000 | | |
| Interview Experience (years) | 4.429 | 2.424 | .376 | 1.827 | .080 | .773 | 1.294 |
| Requirements Experience (years) | 2.868 | 2.542 | .244 | 1.128 | .270 | .698 | 1.432 |
| Development Experience (years) | -.619 | 2.050 | -.071 | -.302 | .765 | .597 | 1.674 |
| Professional Experience (years) | -2.556 | 1.383 | -.421 | -1.848 | .077 | .631 | 1.585 |

Dependent Variable: Adjusted Effectiveness (%)  $R^2$ = .214; Real N = 29; Required N = 82

## 4.4 Results

The results of MLR used to determine the effect of subject experience on the elicitation process effectiveness are shown in TABLE 12 and TABLE 13 respectively.

### 4.4.1 Effect of unfamiliar problem domain experience

As TABLE 12 shows, interview experience ($B_1$=.532), requirements experience ($B_2$=.271) and development experience ($B_3$=.231) tend to have a **positive effect** on effectiveness. On the other hand, professional experience ($B_4$=-.540) has a **negative effect**. According to the interpretation of the non-standardized coefficients (B) established in Section 3.9, the effects in all cases are **zero**. Additionally, none of the effects are **significant** (p-value > 0.05).

Apart from the observed effects, two issues related to the MLR model should be stressed:
1. The coefficient of determination $R^2$ (degree of fit) is very low ($R^2$ = .020 = 2%).
2. The model is not significant (p-value = .860).

This means that the **independent variables (the different types of experience) bear hardly any or no relationship at all to analyst effectiveness.**

### 4.4.2 Effect of familiar problem domain experience

The regression model results are shown in TABLE 12 We find that *interview experience* ($B_1$=4.429) has **a strong positive effect**, whereas professional experience ($B_4$=-2.556), on the other hand, has a **moderate negative effect.** Both cases are very close to statistical significance (p-value = .080 and p-value = .077, respectively), which is notable in view of the limited number of data available (n = 29) for this domain.

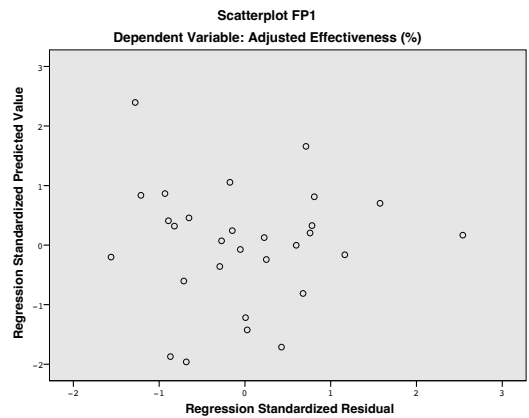

Fig. 3. Scatter plot composed of FP1 model residuals

*Requirements experience* ($B_2$=2.868) and *development experience* ($B_3$=-.619) tend to have a positive and negative effect, respectively, although neither is statistically significant. We believe that requirements experience could have some influence on analyst effectiveness, as the p-value is rather low (p-value = .270) and the model is not statistically powerful. However, this is just a hypothesis.

The coefficient of determination is $R^2$ = .214. The model is not significant (p-value = .199). This was only to be expected in view of the data limitations. We believe that these results should be construed as follows: 1) *interview, requirements* and *professional* experience is related to analyst effectiveness, but 2) **other variables apart from experience are playing a role**.

# 5 DISCUSSION

## 5.1 General Observations

Our results show that for **unfamiliar domains** different experience types (interview, requirements, development and professional) do not have any effect on analyst effectiveness.

For **familiar domains**, the effect varies depending on experience type. **Interview experience clearly has a positive effect**, very close to statistical significance. Requirements experience might also have a positive effect, although our results are nowhere near significant. Note that the size of the sample that we managed to recruit for the familiar domain falls well short of recommendations. Therefore, non-significance could be due to a problem of statistical power. Development experience does not have any effect on analyst effectiveness.

On the other hand, **professional experience has a negative effect**, which is nearly significant, in familiar domains. We did not altogether expect this result. In view of the relative non-specificity of software engineer profiles, we thought there would be positive synergies between all (or many) activity types (e.g., years spent on programming or marketing could provide skills that improve requirements analysis). Our results show that this is not the case, illustrating that professional experience is indeed specific [20]. The same applies to development experience.

Note that the above results are **due exclusively to subject experience**, and domain knowledge is not having any impact. By separately analysing the two domains, we ensure that experience is the only variable influencing analyst effectiveness, as **domain knowledge is blocked**.

As regards knowledge effects, TABLE 12, reporting the results of the MRL for the unfamiliar domain, shows an intercept of 28.73%, whereas TABLE 13, for the familiar domain, has an intercept of 29.25%. The associated standard deviations (2.54% and 5.82%, respectively) suggest that the differences are not substantial, although analysts do, in any case, appear to be slightly more effective in the familiar domain (irrespective of experience). This suggests that domain knowledge has a positive (albeit small) effect on analyst effectiveness, as we reported in all papers [48]. In our view, the real effect is probably greater because: (1) experimental subjects have only very brief contact with FP1, that is, subjects are familiar with the domain concepts but did not work on them for a long time, and (2) subjects found some of the concepts, processes and requirements in FP1 to be particularly alien or hard to find. This means that, if such item types had not been present, subject effectiveness would have been much greater (higher intercepts and, probably, Bs).

## 5.2 Comparison with Related Empirical Work

Our findings are very much consistent with studies by Pitts & Browne [9], Marakas and Elam [4] and Agarwal and Tanniru [6], who, like us, used interviews as a requirements elicitation technique in their research. All three studies conclude that there are no significant differences between novice and experienced subjects, although experienced subjects turned out to be better (albeit very slightly) than novices. However, our results are contrary to findings reported by Niknafs and Berry [5] [17].

The study by Pitts & Browne [9] is an experiment that finds the relationship between development experience and the number of requirements identified by analysts, where the identified effect of r = 0.075. This value is, according to Cohen [54], equivalent to a very small effect. In our case, this correlation is r=-0.06[3], again a very small value, which is, for all practical purposes, negligible. The correlations that we and Pitts and Browne reported are not significant (p-value = 0.591 and 0.661, respectively).

The experimental study by Marakas and Elam [4] reported the effect of experience as a between-group difference (low vs high experience). As a percentage, experienced subjects are 3.09% more effective than the inexperienced subjects. The result is not statistically significant (p-value=0.749). We understand that analysis experience is equivalent to requirements experience. Note that, in order to compare this effect with the findings of our study, we should divide 3.09% by the years of experience of the subjects in the high experience group. This data item is not available in [4]; it only indicates that "*highly experienced subjects were [...] employed at several major corporations*". Assuming that average experience is around 5 to 10 years, the effect per year would be from 3.09% / 5 = .618% to 3.09% / 10 = .309%.

An additional problem with Marakas and Elam's study by that they do not clearly state whether their experiment addresses development or requirements experience. In our case, the effect of development experience is -0.619% (p-value = 0.765), whereas the effect of requirements experience is 2.9 % (p-value = 0.270). The effect of experience that we calculated for Marakas and Elam's study falls in-between our values and is closer to the effect of development experience than requirements experience. This is, in our opinion, a predictable result, as we think that it is easier to recruit subjects with development experience than with requirements experience to perform an experiment.

The study by Agarwal and Tanniru [6] is also experimental. The experimental subjects were students and professionals with at least three years' experience in systems analysis (which, as far as we are concerned, is equivalent to requirements experience). Experienced subjects were slightly more effective (by about 9%) than inexperienced subjects. This result is not significant (the authors state neither size nor direction). It is impossible to compare our results with theirs, as they do not define a gold standard for the rules (their dependent variable) within the experimental problem. We cannot transform their effect of experience into a value that is comparable with ours (percentage of elicited over total items).

Niknafs and Berry [5] found that professional experience had a statistically significant negative effect, whereas requirements experience had a non-significant (with an unspecified size and direction) on relevant and feasible idea generation. For professional experience, Niknafs and

---

[3] Calculated using effect size calculators (see http://www.uccs.edu/~lbecker) with df= N-4 and t taken from the regression model.

Berry [17] again detected a somewhat positive, albeit statistically non-significant, effect with respect to general, relevant and innovative idea generation, and a near significant positive effect for feasible idea generation. For requirements experience, we find that there is a statistically significant negative effect for one of the dependent variables studied (relevant idea generation). With regard to the other variables, the effect of requirements experience is sometimes positive and sometimes negative and is never significant.

Our results are not directly comparable with Niknafs and Berry's findings because the elicitation technique that they apply is visual brainstorming. Some of the subjects that participated in the experiment were familiar and others were unfamiliar with the problem domain. Additionally, the range of requirements and professional experience is confined to from 0 to 4 years, whereas the range of experience is wider (from 0 to 30 years) in our case. Restricting our data to the early years of experience (0 to 4 years) and analysing both domains together using the same analysis technique as used by Nicknafs and Barry, the results are not consistent. One possible reason behind the differences is the elicitation technique used. Brainstorming is an essentially creative process; as they are not influenced by previous experience, subjects that are unfamiliar with the domain are free to suggest ideas. As mentioned in previous studies [48], Niknafs and Berry are, in our opinion, reporting an *einstellung* or functional fixation phenomenon [55] in the field of requirements elicitation.

### 5.3 Differential Effect of Experience by Item Type

Effectiveness could be affected by the manner in which analysts identify the items that define the problem domain (requirements, concepts and processes). For example, the types of items identified by experienced subjects and novices may differ. It seems reasonable to analyse whether experience effects differ by item type.

In order to determine whether the analysts capture one or other item type depending on their experience, we used regression models to study the relationship between experience and analyst effectiveness for each of the items defining the problem domains. Appendix H shows the calculations of the regression models equivalent to TABLE 12 and TABLE 13 separated by concepts, processes and requirements. The results suggest that:

- For the unfamiliar domain, none of the experience types have an effect when analysed separately by processes, concepts and requirements.
- For the familiar domain, interview and requirements experience have positive effects across the board, that is, analysts with more years of interview and requirements experience are better able to identify more processes, concepts and requirements. This is especially true of processes and requirements for which some of the effects are even significant. Note, however, that the effect of development experience tends to be very low, having, for practical purposes, no effect whatsoever, whereas the effect of professional experience is usually negative.

Summarizing: the effects of the experience **do not change** regardless we use the Effectiveness variable, or the number of items of each type (concepts, processes, requirements).

### 5.4 A Possible Reinterpretation of the Experience Effect

The results suggest that, in unfamiliar domains, experience does not improve analyst effectiveness. However, this only applies on average. Fig. 4 shows the two scatter plots for the both unfamiliar domain (a) and the familiar domain (b) depending on years of requirements experience. The hand-drawn red line indicates the minimum analyst effectiveness depending on their experience. Analysts with more years of experience achieve higher minimum effectiveness levels than analysts with less experience

The red line that we have plotted is similar to the *learning curves* often represented in the experience-related literature (e.g., [30]). The learning curves are imaginary lines that denote how much an individual's performance improves as his or her experience increases. In our case, this performance improvement is related not to the individual effectiveness of each subject (effectiveness varies widely depending on years of experience) but to a minimum level of effectiveness that analysts almost always achieve (there is only one exception in FP1). In our opinion, Fig. 4 shows that requirements-related activities improve professional performance more or less as specified by Sim et al. [32]. However, the lack of specialized training, explicit effectiveness evaluation, feedback, etc., prevents performance from improving further.

Practice does not have a positive effect in the case of professional experience, as we discovered in Fig. 5. For the unfamiliar domain (a), the plotted learning curve is much flatter than the curves Fig. 4, as well as being a much worse fit for minimum effectiveness. For the familiar domain (b), this curve simply does not exist. We think that Fig. 5 illustrates that experience is specific, and skills acquired as a result of professional practice cannot be transferred to a specific field like requirements elicitation, as mentioned in Section 5.1.

## 6 VALIDITY THREATS

We grouped the validity threats from two viewpoints: a) threats specifically derived from quasi-experiment execution or contextual aspects, and b) general threats associated with statistical conclusion validity, internal validity, construct validity and external validity.

### 6.1 Specific Validity Threats

The results of our research could be affected by the threats specified below. After checking for each of the threats, we believe that they did not materialize:

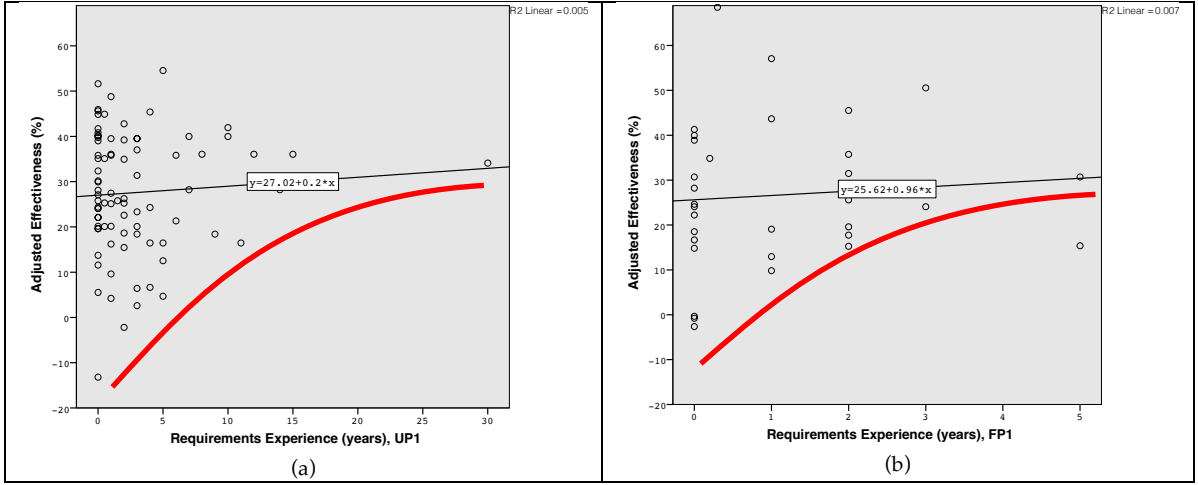

Fig. 4. Minimum effectiveness depending on years of requirements experience

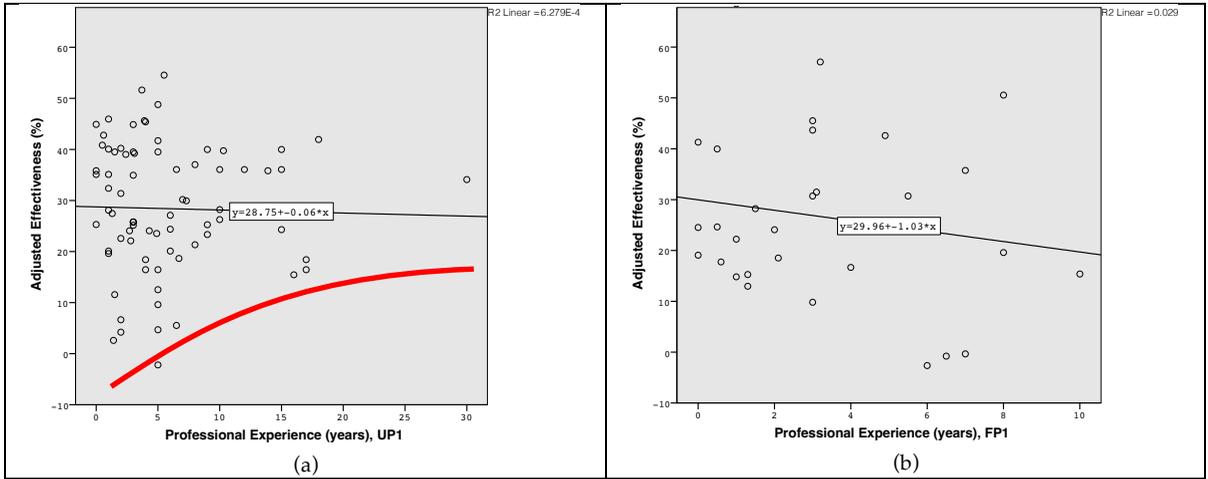

*Fig. 5.* Minimum effectiveness depending on years of professional experience

- *Elicitation time.* The elicitation time available in the elicitation sessions is likely to be insufficient. To check out this possible threat, we studied the relationship between elicitation time and analyst effectiveness. Our results show that they are not correlated. For further details, see Appendix G.1.
- *Number of elicitation sessions.* The failure to detect significant effects for experience may be due to the fact that only one interview was conducted in each of the long series of quasi-experiments that we have conducted. Analyst effectiveness is likely to improve as the number of elicitation sessions increases. Our results show that an increase in the number of sessions (to two) does not result in any improvement in analyst effectiveness. For further details, see Appendix G.2.

Notice that we do not claim that 30 minutes or one elicitation session are enough to perform requirements elicitation *in all cases*. We simply point out that the complexity of the experimental objects was sufficiently small as to subjects become satisfied with their understanding after one ~30-minute session. This impression was corroborated in the post-experimental questionnaire. In turn, most of the subjects manifested that they would perform more interviews if they could.

- *Comparison of individual vs. group interviews.* In group sessions, analyst effectiveness could possibly be affected by knowledge transfer from more experienced to less experienced subjects. As illustrated in Appendix G.3, no such knowledge transfer is observed.
- *Joint analysis of quasi-experiments.* In this study, we have analysed the data jointly without taking into account that they are taken from several experiments. The results of our analysis would not be reliable if the populations of the different quasi-experiments had different characteristics. We analysed the distribution of the model residuals depending on the quasi-experiment to which they belong. This analysis is available in Appendix I. The means and variances of the residuals can be observed to be consistent from one quasi-experiment to another. Additionally, the residuals are scattered around zero. We would expect this to be the case if the populations of the quasi-experiments do not have any influence on analyst effectiveness.

## 6.2 General Validity Threats

The possible threats [56] that we think may affect the quasi-experiments are as follows:

### 6.2.1 Statistical Conclusion Validity Threats

- *Low statistical power:* A total of 90 and 40 experimental subjects are required in both the unfamiliar and familiar problem domains, respectively, to study experience. We managed to gather information from 69 and 29 subjects, respectively; the sample size is therefore insufficient. However, we should note that:
  a) For the unfamiliar domain, the sample size is very close to the required size. In fact, it is greater than necessary to achieve a power of 80% and detect medium-sized effects with $\alpha = 0.10$. Therefore, we believe that, although the results are not significant, they are genuine and not the product of type-II error.
  b) For the familiar domain, interview experience and professional experience have very low p-values (significant at $\alpha = 0.10$, if you prefer). Acting proactively, we considered the independent variables to be influential, anticipating a more than likely type-II error.
- *Violated assumptions of statistical tests.* In the familiar domain, we cannot assure homogeneity of variances. If the variances were not homogeneous, one of the conditions for applying MLR would be violated. Consequently, the results of the inference tests (p-values) could be mistaken. The non-standardized effects (B) would not be affected.
- *Unreliability of measures.* This threat to validity can operate in both the dependent and independent variables:
  a) The dependent variable effectiveness was measured in different settings (for example, individual or group interviews). We assume that some effects derived from the setting could be influencing effectiveness. These effects were eliminated by means of a data adjustment procedure.
  b) The different types of experiences were reported by the experimental subjects using a post-experimental questionnaire. Self-reported values can be unreliable, e.g., inflated, as already observed in [48]. Inflated experience values would flatten the regression lines, leading to low correlations. This threat cannot be counteracted easily, because there is not any reasonable alternative procedure to collect subject experience data. **In our opinion, this is the most serious threat that our research is experiencing.**
- *Measurement bias*. The measurements were made by a single researcher. We cannot rule out that there may have been some bias. For example, subjects studying the unfamiliar domain may have been unintentionally penalized. To prevent this bias, researchers did not access the demographic data until the effectiveness measurement was complete.
- *Restriction of range*. It is not easy to recruit experienced subjects in empirical studies. Our study is no exception. Most subjects had about 0 to 5 years of experience. Only a fraction of the subjects had longer experience. This could have a negative effect on the estimation of effect sizes. However, we believe that this is a secondary problem:
  a) On one hand, most subjects (including the more experienced ones) were interviewed about unfamiliar domains. There is quite a wide range of experience in this domain, on which ground we believe that the result of the MLR is quite reliable. In fact, the MLR with bootstrapping (to overcome the range limitations) yields similar results to the ordinary MLR. In the worst case, if interview and requirements experience had a positive effect on analyst effectiveness, the size of the effect would be low.
  b) In the familiar domain, this threat should tend to support the conclusions of this study, that is, the positive effects of interview experience (and probably requirements experience) and the negative effects of professional experience.

### 6.2.2 Internal Validity Threats

- *Selection*. It can operate in two different ways:
  a) The subjects are not taken from a random sample in any of the quasi-experiments. Therefore, the results may be biased depending on particular characteristics of the populations used. We believe that this threat is unlikely to materialize as multiple quasi-experiments were conducted over several years in at least two settings: academia (UPM) and a professional congress (REFSQ). There is unlikely to be a common moderator variable in all cases.
  b) A possible exception to the former is the *Familiarity* with the problem domain. We assume that the subjects are familiar with FP and unfamiliar with UP1. The former is likely, the latter not necessarily. In the post-experimental questionnaire, we asked subjects about their familiarity with the problem domain. The analysis of the relationship between familiarity and effectiveness yielded non-significant results.
  c) Subjects may be familiar with the problem domains to different degrees. We also questioned the subjects about the problem domain *Difficulty*. Again, we did not identify any clear pattern.
- *Ambiguous temporal precedence*. The conclusions of this research do not draw any causal relations because they are based on quasi-experiments. Even though we studied several experience-related independent variables with respect to experimental task performance, there could be moderator variables that we have not accounted for that could explain the results. As a mitigation strategy, we measured all the observable moderator variables, like warming up, for example, which we considered explicitly in the analysis. We count on other variables, for example, any related to soft skills or analyst personality, offsetting each other in the overall analysis.

### 6.2.3 Construct Validity Threats

- *Construct confounding.* During the execution of the quasi-experiments, we purposely confounded several variables so as not to have to study their interaction:
  a) *Confounding of elicitation time and interview type.* In the quasi-experiments Q-2011 and Q-2012, the increased effectiveness of the subjects could be due to a longer elicitation time (up to 60 minutes) or interview type (group interview) where all the participants have access to exactly the same information. In this case, we cannot separate the effects of the two variables. However, as they were confounded throughout, and neither the interview type nor the elicitation type are key variables in our research, their confounding does not pose a threat to our conclusions.
  b) *Confounding of interviewee and language.* From E-2012 onwards, the interviewee and language variables are confounded. As above, neither the language nor the interviewee are key variables, on which ground their confounding is not a threat.
- *Mono-operation bias.* There are two threats of this type:
  a) The independent variables (the different experiences) were measured in years. Year-based measurement may not accurately represent the level of subject expertise. Other measures (e.g., number of projects) may perhaps have been a better option. Although this may well be the case, research into experience (see Section 2) has historically applied year-based measurement. Additionally, the use of other types of measures does not appear to make a big difference to the results of measurement in years [57]. In our particular case, we analysed a subset of high performing vs. low performing subjects (14 subjects in each group) on a number of variables (including the number of projects), as shown in Appendix G.4. None of the variables was able to explain the differences in subject effectiveness in the unfamiliar domain. In the familiar domain, the number of job positions yields low p-values (< .10), corroborating our results regarding the positive effects of interview experience. Also, in the familiar domain, low p-values (< .10) are associated to subject education; this suggests that personal characteristics may underlie the differences in analyst effectiveness.
  b) The dependent variable (effectiveness) was operationalized in two different ways: 1) as the percentage of items acquired by subjects, and 2) as the total number of items separated by type (requirements, concepts and processes). The effect of experience was similar in both cases.

### 6.2.4 External Validity Threats

- *Interaction of the causal relationship with units.* The fact that experimental subjects are taken from a convenience sample rather than a random sample (that is, subjects are students enrolled in a particular course and not recruited from a larger population) poses a threat to the external validity of the experiment. Therefore, due care should be taken when generalizing our results to professional analysts. However, this threat was addressed since the students were taking a professional master's degree and most had professional experience in computer-related tasks. Therefore, subjects can be considered to be reasonably representative of the developer/analyst population. We believe that our results can at least be generalized to junior developers new to elicitation.
- *Interaction of the causal relationship with settings.* Quasi-experiments were conducted in the laboratory. This means that the setting is completely different to what analysts are used to in the professional world:
  a) The client participating in the interview is a simulation and will not, therefore, act exactly like a real client. To mitigate this threat, the subjects playing the role of interviewee carefully studied the problem domains in order to answer the questions as naturally and realistically as possible.
  b) The software system/s to be developd are not real. To mitigate this threat, the systems used in the quasi-experiments are based on real software systems. We checked that the complexity of the problems was as similar as possible to ensure that problem size is not another factor influencing analyst effectiveness.
  c) The time taken to complete the interviews is limited. We believe that elicitation time was not an obstacle to gathering information. Most subjects finished the elicitation session before the end of the specified time. Additionally, most of the complexity of the original system was eliminated from the experimental objects (that is, the description of the familiar and unfamiliar domains) to assure that they were easy to understand in a short period of time. Finally, there are not statistically significant differences in effectiveness between subjects that considered that the elicitation time was sufficient vs. those who did not.

## 7 CONCLUSIONS

This research studied the effects of different types of experience (interviews, requirements, development and professional) on the elicitation effectiveness of a set of 98 experimental subjects with different levels of experience. Thanks to the experimental design used, we can study the effect of experience separately from the effect of domain knowledge. To do this, we used an unfamiliar and a familiar problem domain, respectively. Our results show that experience has differential effects on analyst effectiveness, depending on the type of problem domain in which the elicitation takes place:

- In the **unfamiliar problem domain, the type of experience** (interviews, requirement, develop-ment or professional) **has no effect whatsoever** on analyst effectiveness. It means that the number of years of professional practice in RE does not substantially improve analyst effectiveness. Fortunately, this only applies on average, as we have found that the **baseline performance increases as analysts acquire interview or requirements experience (**in both unfamiliar and familiar domains). Our interpretation of this result is that, although the RE discipline provides knowledge and strategies that analysts employ by during the elicitation process, it is not sufficient to achieve expertise. A large share of analyst effectiveness is explained by other issues, probably psychological or personal. **Identifying soft skills associated with highly effective analysts** would not only lead to a better understanding of what makes a highly effective analyst but also help to improve analyst training.

- In the **familiar problem domain, interview experience has a** (near significant) **positive effect**. **Professional** experience has a **moderate** (also near significant) **negative effect**. Requirements experience **could have a moderate positive effect**, although the results are nowhere near statistically significant. In our opinion, such results imply that analyst effectiveness is contingent on their previous exposure to the problem domain (and, probably, similar domains as well). The logical consequence is that **requirements analysts should move among development projects in the same (/related) domain(s) exclusively**. Likewise, special **attention should be paid to problem domains in requirements courses**. It is even possible that a substantial share of the training should be spent in studying and carrying out course projects in relevant domains, e.g., banking, insurance, manufacture, health, etc.

Finally, our research shows that analyst effectiveness only improves with specialized training or practice. In other words, development experience and professional experience do not improve analyst effectiveness, but specialized experience (interview and, possibly, requirements experience) does lead to improvement. Therefore, experience behaves similarly in the field of RE to how it does in other branches of knowledge.

The confirmation of previous research in different contexts, and using different methodologies, increases the confidence in the results. However, empirical research is subjected to a large number of threats to validity. Our final word is a call for further research. The communication among stakeholders is a foundation process in requirements engineering, and it is not likely that it disappears or loses relevance in the near future. Elicitation is one the main sources of software quality problems. A better understanding of how elicitation works will contribute to increasing software quality.

## ACKNOWLEDGMENT

This work was supported in part by the project **TIN2011-23216** funded by the Spanish Ministry of Ministry of Economy and Competitiveness. Alejandrina Aranda holds a PhD grant from Itaipú Binacional, Paraguay.

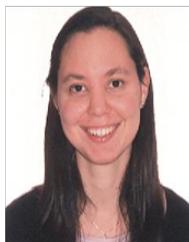

**Alejandrina M. Aranda** received the MSc degree in Information Technology and the PhD degree from the Universidad Politécnica de Madrid. She is a researcher with the UPM's School of Computing. Her research interests include Requirements Engineering and Empirical Software Engineering.

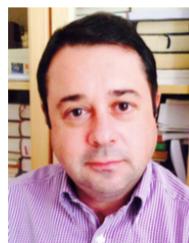

**Oscar Dieste** received his BS and MS in Computing from the University of La Coruña and his PhD from the University of Castilla-La Mancha. He is a researcher with the UPM's School of Computer Engineering. He was previously with the University of Colorado at Colorado Springs (as a Fulbright scholar), the Complutense University of Madrid, and the Alfonso X el Sabio University. His research interests include Empirical Software Engineering and Requirements Engineering.

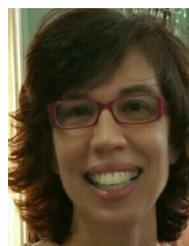

**Natalia Juristo** is *full professor* of software engineering with the Computing School at the Technical University of Madrid (UPM) since 1997 and holds a FiDiPro (Finland Distinguish Professor) research grant since 2013.
She was the Director of the UPM MSc in Software Engineering from 1992 to 2002 and the coordinator of the Erasmus Mundus European Master on SE (whith the participation of the University of Bolzano, the University of Kaiserslautern and the University of Blekinge) from 2007 to 2012.Natalia has served in several *Program Committees* ICSE, RE, REFSQ, ESEM, ISESE and others. She has been *Program Chair* EASE13, ISESE04 and SEKE97 and General Chair for ESEM07, SNPD02 and SEKE01. She has been member of several Editorial Boards, including Transactions on SE, Journal of Empirial Software Engineering and Software magazine. Dr. Juristo has been Guest Editor of special issues in several journals, including Journal of Empirial Software Engineering, IEEE Software, Journal of Software and Systems, Data and Knowledge Engineering and the International Journal of Software Engineering and Knowledge Engineering.

# APPENDIX A: QUESTIONNAIRE

## Academic Education

What bachelor's degree are you studying for?

How many years of study does it take to earn your degree?

Select the requirements engineering topics that you studied before enrolling for this UPM master's programme;
- ( ) Elicitation techniques
- ( ) Natural language analysis
- ( ) Data modelling
- ( ) Process modelling
- ( ) Use cases / user stories
- ( ) Requirements writing
- ( ) Requirements tools
- ( ) Reviews
- ( ) Mock-ups / storyboards / prototyping
- ( ) Change management
- ( ) Formal specification
- ( ) Model checking
- ( ) Product lines
- ( ) Others

If you have checked "others", please specify further:

Have you ever attended a course or seminar on requirements elicitation?
- Never
- As part of my bachelor's programme
- A specialized seminar

Have you completed assignments or practical exercises on requirements elicitation?
- Never
- As part of my bachelor's programme
- A specialized seminar

## Professional Experience

Are you currently:
- Studying?
- Working?
- Studying and working?

If you are working, specify your position:
Specify previous positions you have held (if any):

Have you participated in a real project as an analyst? How many times?

How many years of interviewing experience do you have? (If you are not sure, give an approximate number; numbers may contain decimals)

How many years of elicitation experience do you have? (If you are not sure, give an approximate number; numbers may contain decimals)

Rate your skills in (from 1= very low and 5 = very high):

| SKILL | 1 | 2 | 3 | 4 | 5 |
|---|---|---|---|---|---|
| Interviews | | | | | |
| Requirements Specification | | | | | |

## Knowledge of the problem domain

Rate your knowledge of the problem stated in the exercises?

| PROBLEM | 1 | 2 | 3 |
|---|---|---|---|
| AP1 | | | |
| IP1 | | | |
| 1. It is the first time you have dealt with this type of problem | | | |
| 2. The problem is familiar to you but you have never worked on it before | | | |
| 3. The problem is familiar to you and you have worked on it before | | | |

Rate the level of difficulty of the problem stated in the exercises? (from 1 = low and 3 = high)

| PROBLEM | 1 | 2 | 3 |
|---|---|---|---|
| AP1 | | | |
| IP1 | | | |

## Elicitation and Conceptualization

Do you think you were given enough time for the elicitation process?
- ( ) Yes
- ( ) No

If not, explain why:

Do you think you were given enough time to write the requirements specification during the exercise?
- ( ) Yes
- ( ) No

If not, explain why:

## Need of further interviews

In the first simulated interview, do you think that you understood the problem domain and identified the main requirements? If you think there were differences among problems, please give a brief explanation:

Do you think that more interactions with the customers (that is, more interviews) are necessary to understand the different problems? How many?

*Thanks for taking this survey.*

## APPENDIX B: PROBLEM DOMAINS

The following is a detailed description of the items (requirements, concepts and processes) constituting the problem domains used in the baseline experiment: FP1 and UP1.

Note that the descriptions of the requirements and models are guidelines. They are designed as a checklist to establish whether subjects identify the problem domain items during the elicitation session. Measurement should by no means be construed as a literal process: rather than applying a blind benchmarking procedure, we use the checklist to interpret which items the subject has identified and reported.

TABLE 1 and Fig. 1 and Fig. 2 describe the requirements, concepts and processes, respectively, for the **Text Messaging** problem **(FP1)**

TABLE 1
ITEMS DEFINING THE FAMILIAR PROBLEM FP1

| TYPE | ITEM ID | DESCRIPTION |
|---|---|---|
| Requirements | R1 | Users will be able to connect to the chat server by entering an ID and password. |
| | R2 | The ID and password will be created via a web page. |
| | R3 | If the ID and password are incorrect, the system will display a message. |
| | R4 | After the system has correctly connected to the server (internal company network), it will wait for user input or reception of a chat message. |
| | R5 | The system will be able to run in both the foreground and background. The system will provide the option to: |
| | R6 | Select the chat/contact in/with which users wish to communicate (Select contact). |
| | R7 | Write a message. |
| | R8 | The application will only send plain text messages. (Images, icons, multimedia messages, etc., will not be included.) There is no limit on the size (number of characters) per message. |
| | R9 | Use a user ID/email address to enter contact. |
| | R10 | Specify user ID to delete contact. |
| | R11 | Accept an entry request. |
| | R12 | Reject an entry request. |
| | R13 | Send a message. |
| | R14 | Messages will be sent between two users only. |
| | R15 | The system will request explicit confirmation from users before sending messages/ requests to server. |
| | R16 | A chat selected by the user will open and display all the unread messages. |
| | R17 | Unread messages will be highlighted in bold. |
| | R18 | The system will display a message log according to the following parameters or options: 1. Last hour or 2. Last four messages. |
| | R19 | If the system is running in the background: A pop-up will notify users that a message has been received (depending on the operating system). |
| | R20 | The message will be added to the respective chat, and an unread messages counter will be updated. |
| | R21 | The notification will be added to the respective chat, and an unread messages counter will be updated. |
| | R22 | The pop-up will be accompanied by a sound. |
| | R23 | If the system is running in the foreground: If the message/notification is for the current chat, the message/notification will be displayed directly. |
| | R24 | The date and time will be displayed together with the message/notification. |
| | R25 | Otherwise, the message/notification will be added to the respective chat, and an unread messages/notifications counter will be updated, BUT the user will not be notified (as in the iPhone). |
| | R26 | For delete notifications, both the input and the contact will be deleted immediately. |
| | R27 | Users will be able to disconnect the chat server. |
| | R28 | When the system receives a message/notification from the server, the following actions will be taken: (only if the program is running; otherwise, the messages will be lost). |
| Concepts | C1 | User |
| | C2 | Contacts |
| | C3 | Messages |
| | C4 | Text |
| | C5 | Chat |
| | C6 | Entry request |
| | C7 | Deletion request |
| | C8 | Entry confirmation |
| | C9 | Entry rejection |
| | C10 | Log |
| Processes | A1 | Connect to server |
| | A2 | Wait for user input |
| | A3 | Request contact entry |

| Type | Item ID | Description |
|---|---|---|
| | A4 | Accept entry request |
| | A5 | Reject entry request |
| | A6 | Request contact deletion |
| | A7 | Select contact (chat) |
| | A8 | Write message |
| | A9 | Send message |
| | A10 | Receive message |
| | A11 | Send request |
| | A12 | Receive request |
| | A13 | Notify message reception |
| | A14 | Execute deletion request |
| | A15 | Close notification |
| | A16 | Display message log (show message in chat) |

**FP1 – Activity Diagram**

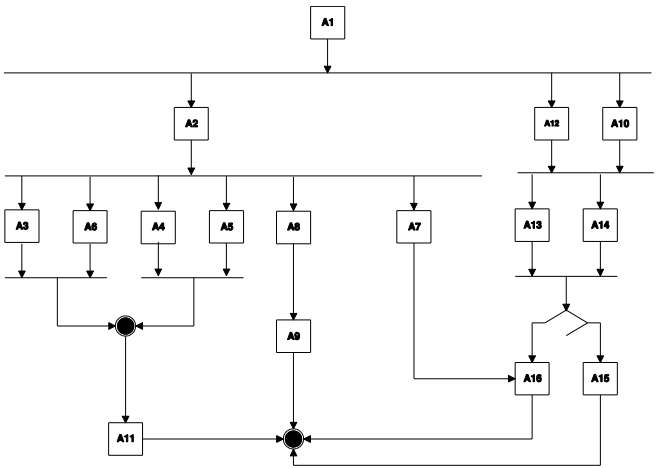

Fig. 1. Activity diagram for familiar problem – baseline experiment

**FP1 - Concept Model**

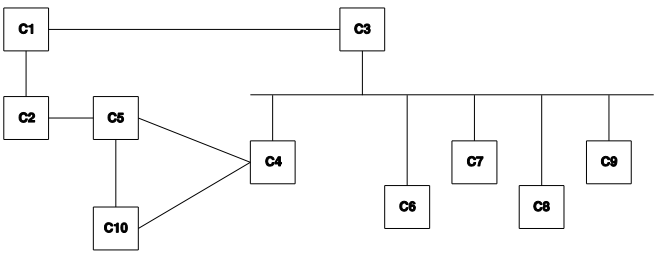

Fig. 2. Model of familiar problem concepts – baseline experiment

TABLE 2 and Figs. 14 and 15 describe the requirements, concepts and processes of the battery recycling problem **(UP1).**

TABLE 2

ITEMS DEFINING THE UNFAMILIAR PROBLEM UP1

| Type | Item ID | Description |
|---|---|---|
| Requirements | R1 | The system will provide an option for manually starting up the separator. |
| | R2 | The system will provide an option for automatically shutting down the separator when any of the three output bins contains a specified quantity of batteries. |
| | R3 | The system will provide an option for manually starting up the shredder. |
| | R4 | The system will provide an option for automatically shutting down the shredder when the output bin contains a specified quantity of battery powder. |
| | R5 | The system will provide an option for manually starting up the crusher. |
| | R6 | The system will provide an option for automatically shutting down the crusher when the output bin contains a specified quantity of battery powder. |
| | R7 | The system will provide an option for manually starting up the distiller. |
| | R8 | The system will provide an option for automatically shutting down the distiller when any of the output bins contains a specified quantity of metal. |
| | R9 | The system will shut down the distiller when the distillation time has expired. |
| | R10 | The system will provide an option for registering the corresponding recycling batch for the ash stored in each waste vat. |
| | R11 | The system will provide an option for registering the delivery notes provided by the battery suppliers. |
| | R12 | The system will provide an option for registering the allocated recycling batch for the batteries on each delivery note. |
| | R13 | The system will provide an option for registering the collection notes provided by the vat transport suppliers. |
| | R14 | The system will provide an option for registering the vats corresponding to each collection note. |
| | R15 | The system will provide an option for registering the quantity of metals extracted by the recycling process. |
| Concepts | C1 | Metal |
| | C2 | Manganese |
| | C3 | Mercury |
| | C4 | Nickel |
| | C5 | Cadmium |
| | C6 | Copper |

| Type | Item ID | Description |
|---|---|---|
| Concepts | C7 | Zinc |
| | C8 | Battery |
| | C9 | General-purpose batteries |
| | C10 | Button cells |
| | C11 | Button cells containing manganese |
| | C12 | Button cells not containing manganese |
| | C13 | Supplier |
| | C14 | Delivery note |
| | C15 | Batch |
| | C16 | Vat |
| | C17 | Ash |
| | C18 | Collection note |
| | C19 | Machine |
| | C20 | Separator |
| | C21 | Shredder |
| | C22 | Distiller |
| | C23 | Incinerator |
| | C24 | Crusher |
| Processes | A1 | Enter delivery note |
| | A2 | Separate batteries |
| | A3 | Crush button cells not containing manganese |
| | A4 | Shred general-purpose batteries |
| | A5 | Crush button cells containing manganese |
| | A6 | Distil with steel |
| | A7 | Distil without steel |
| | A8 | Precipitate |
| | A9 | Send message |
| | A10 | Receive message |
| | A11 | Send request |
| | A12 | Receive request |
| | A13 | Notify message reception |
| | A14 | Execute deletion request |
| | A15 | Close notification |
| | A16 | Display message log (show message in chat) |

**UP1 – Activity Diagram**

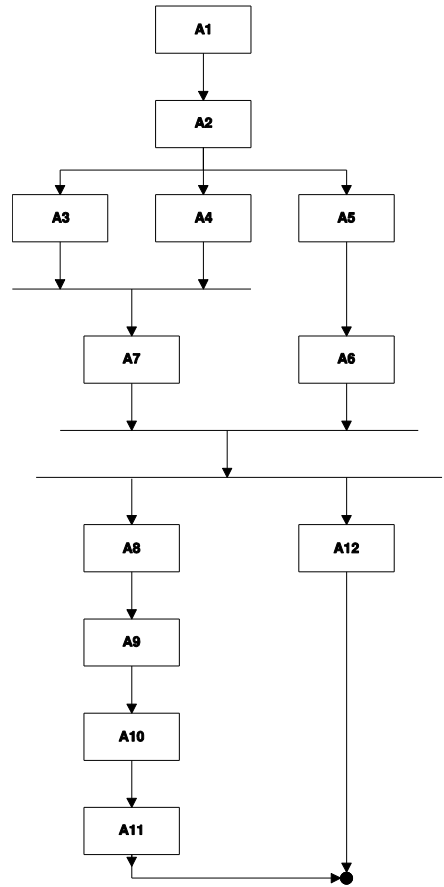

Fig. 3. Activity diagram for unfamiliar problem — baseline experiment

**UP1 - Concept Model**

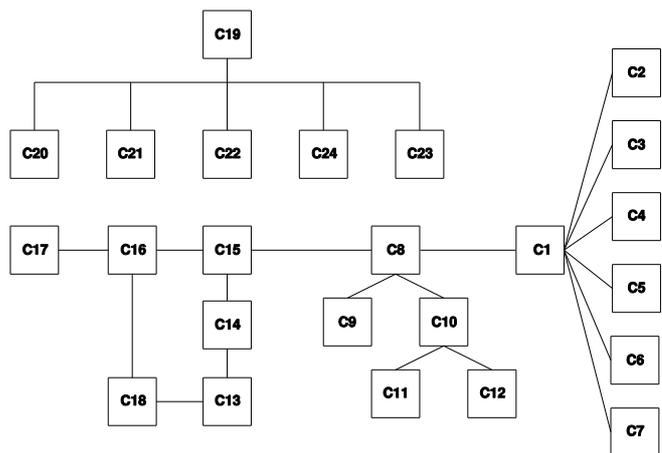

Fig. 4. Concept model of unfamiliar problem — baseline experiment

## APPENDIX C: DISTRIBUTION OF EXPERIENCE

### C.1- UNFAMILIAR DOMAIN

Fig. 5 shows the distribution of the different experience types under study for the unfamiliar problem (UP1) measured in years. As the bar charts show, the range of years of requirements and professional experience is quite wide (from 0 to 30). On average, subjects claim to have 4.5 years of requirements experience (Fig. 5.b) and 5.9 years of professional experience (Fig. 5.d). As regards the years of interview and development experience, the spread is smaller. On average, subjects have 4.5 years of requirements experience (Fig. 5.a) and 4 years of development experience (Fig. 5.b).

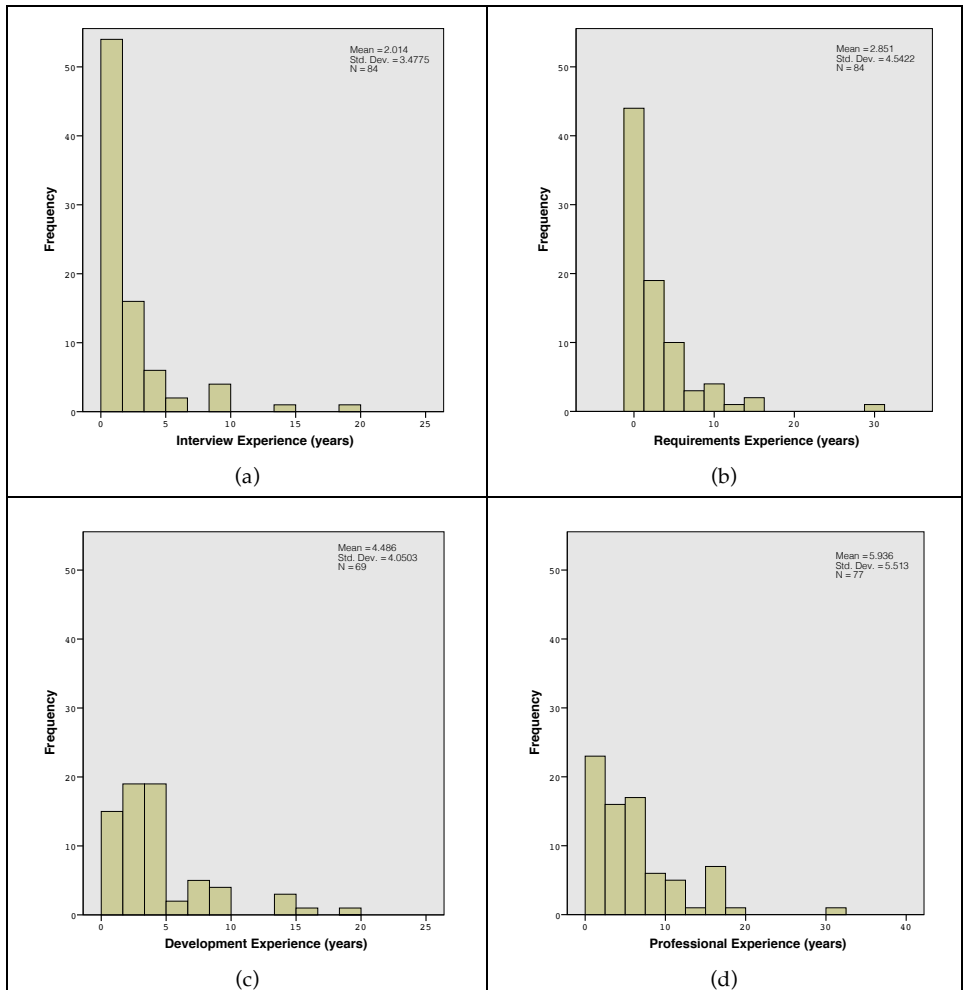

Fig. 5. Distribution of a) interview, b) requirements, c) development and d) professional experience, UP1

### C.2- FAMILIAR DOMAIN

Fig. 6 shows the distribution in years of the different types of experience under study for the familiar problem (FP1). As the bar charts show, the distribution of years of interview and requirements experience ranges from 0 to 6 and 0 to 5 years, respectively. The subjects have very little interview (Fig. 6.a) and requirements (Fig. 6.b) experience, amounting to on average one year. As regards the years of development and professional experience, the range is wider. On average, subjects have 2 years of development experience (Fig. 6.c) and 2.7 years of professional experience (Fig. 6.d).

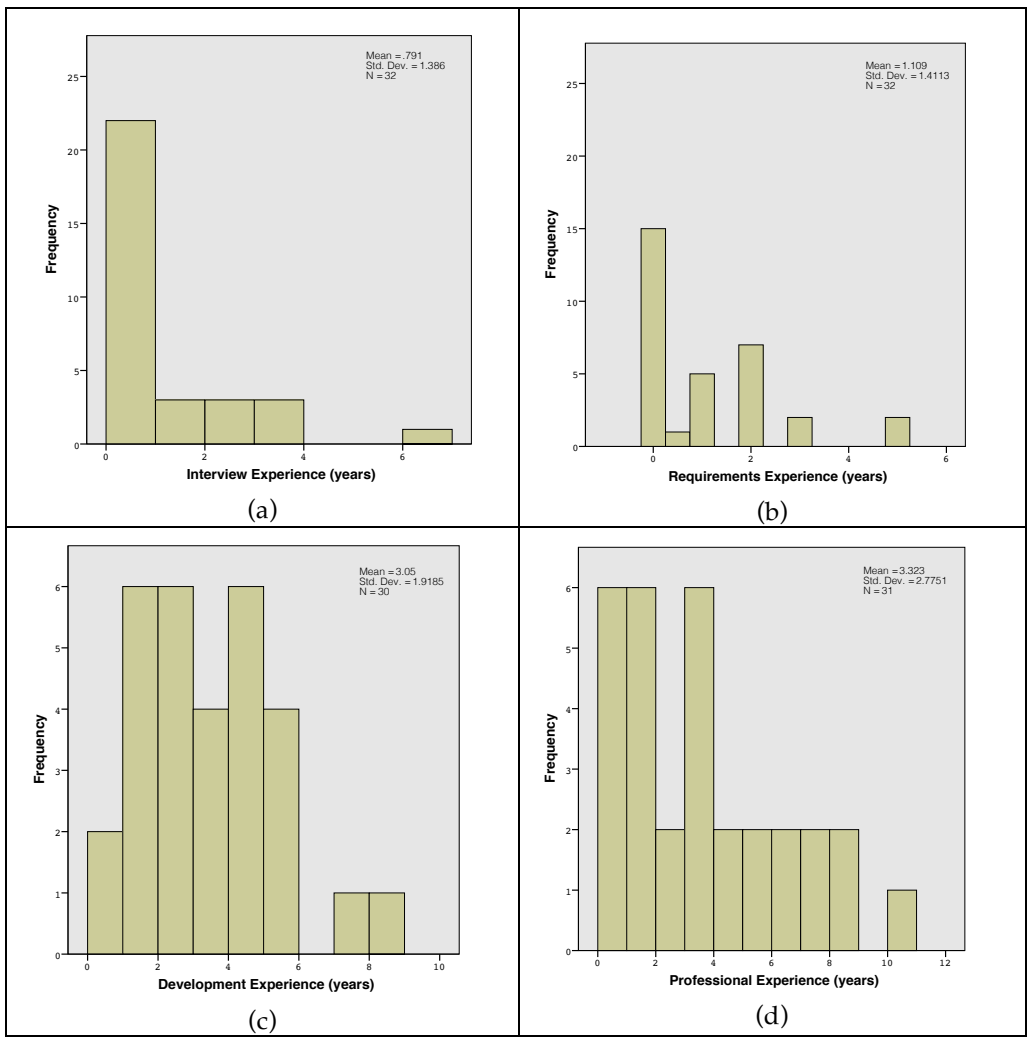

Fig. 6. Distribution of a) interview, b) requirements, c) development and d) professional experience, FP1

# APPENDIX D: Data Adjustment

## D.1- Adjustment Procedure: an example

By way of an example, we present the procedure for calculating the percentage of adjustment for the **Execution Effect**. Note that the execution effect accounts for whether or not subjects participated in the experiment before or after taking the Requirements Engineering course at UPM. To calculate this effect, we used the weighted difference between total average effectiveness achieved by subjects in the experiments that were run before teaching RE and the total average effectiveness achieved in the experiments that were conducted at the end of the course, where the other moderator variables are constant.

Looking at the moderator variables that influence the experiments Q-2007 and E-2012 in TABLE 3, for example, we find that that they all match, except for Execution. Therefore, the execution effect could be calculated as:

$$ExecutionEffect = |Effectiveness\ Q\text{-}2007 - Effectiveness\ E\text{-}2012| = |39\text{-}25| = |14| = 14$$

Note that, as it is still an estimation, we calculate the execution effect as an integer value.

It would not be possible to calculate the execution effect using the experiments Q-2012 and E-2012, since, as TABLE 4 shows, the Execution variable is the same for both experiments and the other moderator variables (like InterviewType and Interviewee) are not constant.

According to this procedure, we receive two possible adjustment values for this case of the execution effect:

$$ExecutionEffect_1 = |Effectiveness\ Q\text{-}2007 - Effectiveness\ E\text{-}2012| = |39\text{-}25| = |14| = 14$$

$$ExecutionEffect_2 = |Effectiveness\ Q\text{-}2012 - Effectiveness\ Q\text{-}2011| = |7 - 48| = |\text{-}11| = 11$$

Each of these adjustment values has been taken from studies that have different sample sizes. Larger studies are usually considered to be more accurate than smaller-sized studies. Therefore, we use the weighted mean (depending on sample size) of individual effects to calculate a single execution effect. Accordingly, we get:

$$ExecutionEffect_a = (14*14 + 11*37)/51 = 11.82 = 12$$

that is, subjects tend, on average, to be 12% more effective after they have received training in RE.

### TABLE 3
Execution effect when the other moderator variables are constant

Consolidation Effectiveness (%)

| Experiment | Interview Type | Execution | Warming Up (Training) | Interviewee | N | Mean |
|---|---|---|---|---|---|---|
| Q-2007 | Individual | After | None | OD | 7 | 39.22 |
| E-2012 | Individual | Before | None | OD | 7 | 24.93 |
| | | | | JWC | 7 | 50.98 |

### TABLE 4
Execution effect when the other moderator variables are not constant

Consolidation Effectiveness (%)

| Experiment | Interview Type | Execution | Warming Up (Training) | Interviewee | N | Mean |
|---|---|---|---|---|---|---|
| Q-2012 | Group | Before | None | OD | 21 | 37.07 |
| E-2012 | Individual | Before | None | OD | 7 | 24.93 |
| | | | | JWC | 7 | 50.98 |

## D.2- Percentage Adjustment: effectiveness

TABLE 5 shows a brief description of each moderator variable, as well as the respective percentage adjustment, which we gathered by means of the procedure specified in the above example. TABLE 6 shows the adjustment procedure applied to the data of each of the quasi-experiments.

TABLE 5

PERCENTAGE ADJUSTMENT FOR EACH MODERATOR VARIABLE

| MODERATOR VARIABLE | DEFINITION | UP1 PERCENTAGE ADJUSTMENT | FP1 PERCENTAGE ADJUSTMENT |
|---|---|---|---|
| Execution | The increase in subject effectiveness caused by the weighted difference between average total effectiveness achieved by subjects in the experiments that were executed before and after subjects received specialized requirements training. | 12% | N/A |
| Interview Type | The increase in subject effectiveness caused by the group with respect to individual interviews. | 11% | N/A |
| Warming Up | Warming up refers to the duration of the short training. In this case, given the contextual differences between experiments, we identified three types of warming up, equivalent to one, two and six weeks, respectively. | | |
| WUpEffect1 | The difference in total average effectiveness between the experiment without warming up (Q-2012) and the experiment with a one-week warming up (Q-2013). | -1% | -6% |
| WUpEffect2 | The difference in total average effectiveness between the experiment without warming up (Q-2012) and the experiment with a two-week warming up (Q-2015). | 7% | 0 |
| WUpEffect6 | The difference in total average effectiveness between the experiment without warming up (Q-2012) and the experiment with a six-week warming up (Q-2014). | 10% | 5% |
| Interviewee | The interviewee effect is equivalent to the additional information provided by one interviewee with respect to another. Three interviewees participated in the sequence of experiments (OD, AG and JWC). | | |
| IntervieweeODAG | The additional information provided by interviewee AG with respect to interviewee OD. | 23% | NA |
| IntervieweeODJWC | The additional information provided by interviewee JWC with respect to interviewee OD. | 18% | 23% |

TABLE 6

ADJUSTMENT PROCEDURE

| ADJUSTMENT OF THE EXPERIMENTAL DATA FOR EACH OF THE QUASI-EXPERIMENTS |
|---|
| Adjusted Q-2007 = Q-2007 Effectiveness – Execution Effect |
| Adjusted Q-2009 = Q-2009 Effectiveness – Execution Effect– IntervieweeODAG Effect |
| Adjusted Q-2011 = Q-2011 Effectiveness – Execution Effect– InterviewType Effect |
| Adjusted Q-2012-REFSQ = Q-2012 Effectiveness – InterviewType Effect |
| |
| OD Adjusted E-2012-UP1 = E-2012_OD Effectiveness |
| JWC Adjusted E-2012-UP1 = E-2012_JWC Effectiveness – IntervieweeODJWC Effect |
| OD Adjusted E-2013 -UP1 = E-2013 OD Effectiveness – WarmingUp Effect1 |
| JWC Adjusted E-2013-UP1 = E-2013 JWC Effectiveness – IntervieweeODJWC Effect - WarmingUp Effect1 |
| OD Adjusted E-2014-UP1 = E-2014 Effectiveness –WarmingUp Effect6 |
| OD Adjusted E-2015-UP1 = E-2015 OD Effectiveness – WarmingUp Effect2 |
| JWC Adjusted E-2015-UP1 = E-2015 JWC Effectiveness – IntervieweeODJWC Effect – WarmingUp Effect2 |
| |
| OD Adjusted E-2012-FP1 = E-2012_OD Effectiveness |
| JWC Adjusted E-2012-FP1 = E-2012_JWC Effectiveness – IntervieweeODJWC Effect |
| OD Adjusted E-2013 -FP1 = E-2013 OD Effectiveness – Warming Up Effect1 |
| JWC Adjusted E-2013-FP1 = E-2013 JWC Effectiveness – IntervieweeODJWC Effect - WarmingUp Effect1 |
| OD Adjusted E-2014-FP1 = E-2014 Effectiveness –Warming Up Effect6 |
| OD Adjusted E-2015-FP1 = E-2015 OD Effectiveness – WarmingUp Effect2 |
| JWC Adjusted E-2015-FP1 = E-2015 JWC Effectiveness – IntervieweeODJWC Effect – WarmingUp Effect2 |

## D.3- ADJUSTMENT PROCEDURE: BY ITEM TYPE

As for the response variable effectiveness, the performance of the subjects classed by item type (processes, concepts and requirements) was adjusted according to the moderator variables: Execution, Interview Type, Interviewee and Warming Up, as shown in TABLE 7. Note that, in this case, the adjustment values are in natural units and not percentages.

TABLE 7

ADJUSTMENT PROCEDURE FOR ITEM TYPE

| DOMAIN | MODERATOR VARIABLE | ADJUSTMENT VALUE | | |
|---|---|---|---|---|
| | | PROCESSES | CONCEPTS | REQUIREMENTS |
| UP1 | Execution | 1 | 2 | 5 |
| | Interview Type | 1 | 3 | 2 |
| | IntervieweeODAG | 1 | 6 | 4 |
| | IntervieweeODJWC | 2 | 6 | 2 |
| | Warming Up | | | |
| | 1 week | 1 | 3 | 2 |
| | 2 weeks | 1 | 3 | 1 |
| | 6 weeks | 1 | 4 | 0 |
| FP1 | Execution | N/A | N/A | N/A |
| | Interview Type | N/A | N/A | N/A |
| | IntervieweeODAG | N/A | N/A | N/A |
| | IntervieweeODJWC | 3 | 3 | 6 |
| | Warming Up | | | |
| | 1 week | 1 | 4 | 1 |
| | 2 weeks | 2 | 0 | 1 |
| | 6 weeks | 0 | 2 | 1 |

## APPENDIX E: Regression Models

### E.1- FP1 Problem Domain

TABLE 8TABLE 10 shows the results of the MLR with the three variables (interview, requirements and professional experience) for the unfamiliar domain. We find that interview experience (B1=.114) and requirements experience (B2=.497) tend to have a positive effect on effectiveness.

On the other hand, professional experience (B3=-.461) has a negative effect. Note that the coefficient of determination $R^2$ (degree of fit) is very low ($R^2$ = .013 = 1%), and the variables are nowhere near statistically significant in any of the cases.

TABLE 8

Experience Effect through the regression model for UP1

| Model | | Non-standardized coefficients | | Typified coefficients | T | Sig. |
|---|---|---|---|---|---|---|
| | | B | Std. Error | Beta | | |
| 1 | (Constant) | 29.423 | 2.378 | | 12.375 | .000 |
| | Interview Experience (years) | .114 | .727 | .031 | .157 | .875 |
| | Requirements Experience (years) | .497 | .629 | .175 | .791 | .431 |
| | Professional Experience (years) | -.461 | .512 | -.191 | -.902 | .370 |

a. Dependent variable: Adjusted effectiveness (%)  $R^2$ = 0.013 N= 76

### E.2- UP1 Problem Domain

TABLE 9 shows the results of the MLR with the three variables (interview, requirements and professional experience) for the familiar domain. We find that interview experience (B1=4.207) and requirements experience (B2=2.675) tend to have a positive effect on effectiveness. On the other hand, professional experience (B3=-2.694) has a negative effect. Note that the coefficient of determination R2 (degree of fit) is very low, albeit non-negligible (R2 = .211 = 21%). The effect of professional experience is statistically significant (p-value = .046). The effect of interview experience is also very close to significance (p-value = 0.075).

TABLE 9

Effect of experience by means of regression model for FP1

| Model | | Non-standardized coefficients | | Typified coefficients | T | Sig. |
|---|---|---|---|---|---|---|
| | | B | Std. Error | Beta | | |
| 1 | (Constant) | 28.288 | 4.790 | | 5.906 | .000 |
| | Interview Experience (years) | 4.207 | 2.267 | .357 | 1.855 | .075 |
| | Requirements Experience (years) | 2.675 | 2.415 | .228 | 1.108 | .279 |
| | Professional Experience (years) | -2.694 | 1.281 | -.444 | -2.104 | .046 |

a. Dependent variable: Adjusted effectiveness (%)  $R^2$ = 0.211 N= 29

## APPENDIX F: COLLINEARITY

### F.1- UP1 PROBLEM DOMAIN

TABLE 10 shows the results of the collinearity diagnosis corresponding to the MLR with four variables for the unfamiliar domain. We find that the condition index (CI= 7.166) is within the expected range (CI < 10, low collinearity). From the viewpoint of the collinearity condition, the model is perfectly valid for studying the effects of experience on the effectiveness of the subjects.

TABLE 10

EFFECTS OF EXPERIENCE – MLR FOR UP1

Collinearity Diagnostics

| | | | | Variance Proportions | | | | |
|---|---|---|---|---|---|---|---|---|
| Model | Dimension | Eigenvalue | Condition Index | (Constant) | Interview Experience (years) | Requirements Experience (years) | Development Experience (years) | Professional Experience (years) |
| 1 | 1 | 3.871 | 1.000 | .02 | .01 | .01 | .01 | .01 |
| | 2 | .674 | 2.396 | .24 | .10 | .03 | .06 | .00 |
| | 3 | .250 | 3.938 | .63 | .11 | .00 | .47 | .01 |
| | 4 | .130 | 5.447 | .01 | .77 | .37 | .29 | .09 |
| | 5 | .075 | 7.166 | .09 | .01 | .59 | .17 | .90 |

a. Dependent Variable: Adjusted Effectiveness (%)

### F.2. FP1 PROBLEM DOMAIN

TABLE 11 shows the results of the collinearity diagnosis corresponding to the MLR for the familiar domain. Note that the condition index (CI= 7.166) is smaller than 10. From the viewpoint of variable collinearity, Model 2 is valid enough for studying the effects of experience on the effectiveness of the subjects.

TABLE 11

COLLINEARITY DIAGNOSIS – MODEL 2

Collinearity Diagnostics

| | | | | Variance Proportions | | | | |
|---|---|---|---|---|---|---|---|---|
| Model | Dimension | Eigenvalue | Condition Index | (Constant) | Interview Experience (years) | Requirements Experience (years) | Development Experience (years) | Professional Experience (years) |
| 1 | 1 | 3.778 | 1.000 | .01 | .02 | .02 | .01 | .01 |
| | 2 | .533 | 2.661 | .05 | .87 | .03 | .00 | .01 |
| | 3 | .364 | 3.223 | .21 | .00 | .75 | .02 | .00 |
| | 4 | .199 | 4.360 | .25 | .02 | .16 | .01 | .90 |
| | 5 | .126 | 5.483 | .48 | .09 | .04 | .96 | .07 |

a. Dependent Variable: Adjusted Effectiveness (%)

## APPENDIX G: INFLUENCE OF OTHER FACTORS ON EFFECTIVENESS

### G.1- EFFECT OF ELICITATION TIME

Fig. 7 shows the scatter plots for both problem domains. The x-axis shows the time taken by subjects to complete the elicitation sessions. The y-axis represents the effectiveness of subjects. Our results, as the plots show, suggest that time has a positive effect with a value of 0.4%-0.5% per minute.

However, the p-values (0.261, 0.396) are nowhere near statistically significant. Note that we have not used the data of the experiments with group interviews (Q-2011 and Q-2012) in Fig. 7, as elicitation time and interview time are confounded.

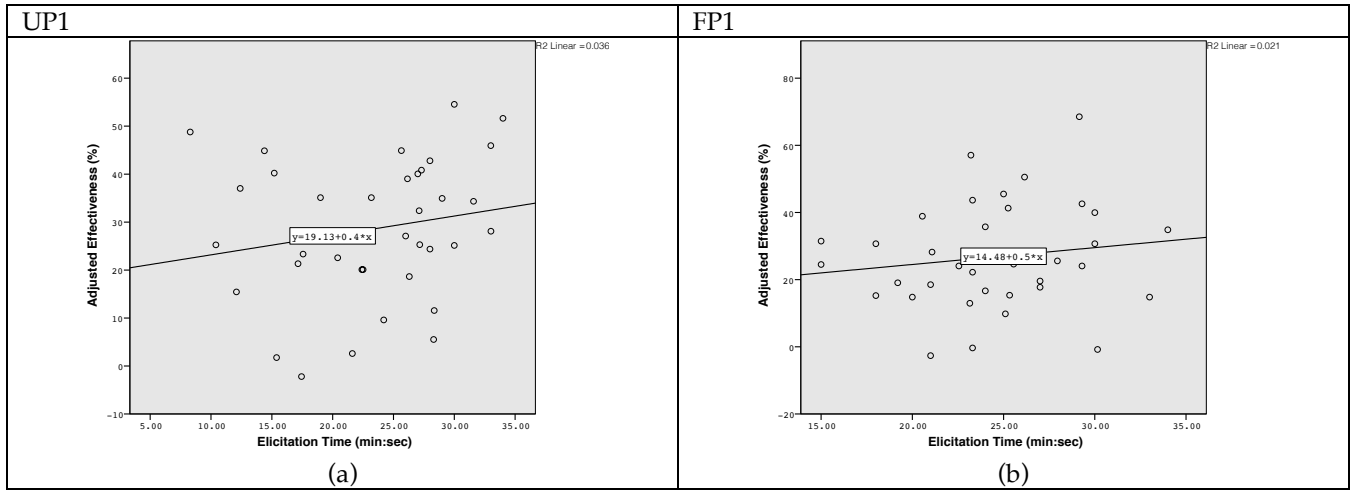

Fig. 7. Relationship between elicitation time and analyst effectiveness

### G.2- EFFECT OF INTERVIEW 2

In Q-2013, the experimental subjects had the chance to hold a second interview with the aim of rounding out the information gathered in the first interview. Fig. 8 shows the profile plots, illustrating, for each problem domain, the effectiveness achieved by the subjects in both the first and second sessions.

As Fig. 8.a shows, the subjects achieved similar effectiveness for the unfamiliar domain. In both the first and second elicitation session, the subjects who interviewed JWC achieved an effectiveness of 44%, whereas the effectiveness of subjects who interviewed OD was 29%. In the familiar problem domain (Fig. 8.b), we found that there was as difference of 3% for JWC and 5% for OD, which, for all practical purposes, is negligible.

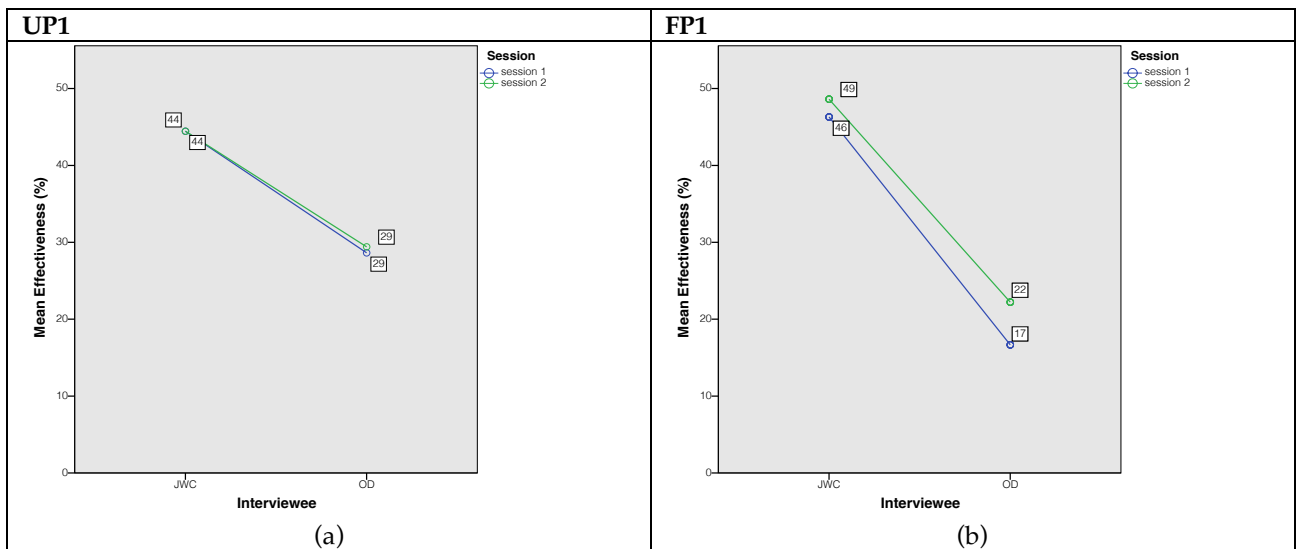

Fig. 8. Effect of Interview 2 classified by problem domain

Therefore, we did not find any improvement in the effectiveness of analysts in either the familiar and unfamiliar domains. These results are not conclusive, and further research is required. One possible question is what happens with analyst effectiveness in follow-on interviews. Are analysts with domain knowledge more effective in proportion to the number of interviews held? Does analyst experience improve their effectiveness in unfamiliar problem domains? Both questions could help to better understand the elicitation activity.

**G.3- INDIVIDUAL VS. GROUP INTERVIEWS**

Six of the series of quasi-experiments (Q-2007, Q-2009 and Q2012) were conducted by means of individual open interview. The other two (Q-2011 and Q-2012-REFSQ) were, due to contextual constraints, carried out by means of group open interviews. Note that only the unfamiliar problem (UP1) was tested in Q-2011 and Q-2012-REFSQ.

The group interview can have an influence on both average and individual subject effectiveness. The major risk is that less experienced subjects benefit from the information arising out of the conversation generated by more experienced subjects, that is, less experienced analyst effectiveness could increase over and above what they would be capable of achieving on their own thanks to the information transfer from more experienced analysts.

This risk is not purely theoretical. In group interviews, we observed that only one subset of subjects asked questions. These subjects were precisely the ones with a better reputation within the group, which is associated with the recognition by the group of their experience, expertise or any other positive quality.

Information transfer from expert to novice analysts can be explored using on the same data that we used in our research. If there were any such transfer, we should observe more or less horizontal regression lines in the group interview (analysts would be more or less equally effective, irrespective of their years of experience), whereas the regression line for individual interviews should be upward (as there is no transfer, expert analysts would be more effective than novices).

We did not use regression models to make this comparison, as the size after dividing the data set into two groups of approximately 40 subjects was nowhere near large enough to fit a four-variable model. Note that any possible group information transfer will apply to the unfamiliar problem domain only, as we did not perform any group interview in the familiar domain.

Fig. 9 shows the regression lines between (a) interview and (b) requirements experience and analyst effectiveness for group interviews. The correlation coefficients are small to medium ($r = 0.27$ and $r = 0.193$, respectively) and not significant ($p\text{-value} = 0.096$ and $p\text{-value} = 0.252$, respectively). The p-value associated with interview experience is almost significant.

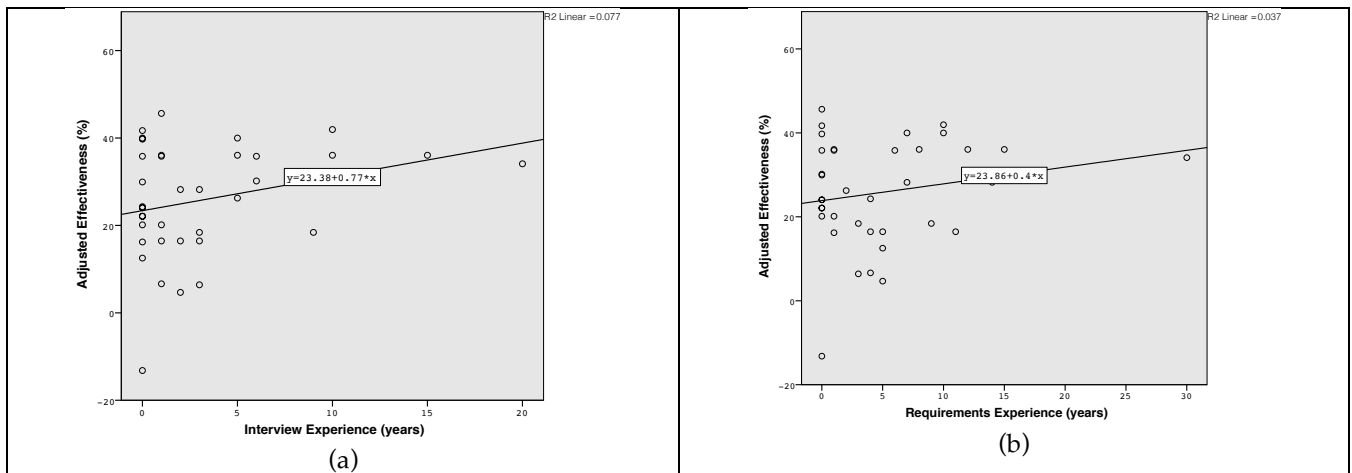

Fig. 9. Regression lines for group interviews

The regression lines are completely contrary to what we would expect to find should there be any information transfer from expert to novice analysts in the group interview. The regression lines of Fig. 9 show upward slopes, which means that the novice analysts behave worse than experienced analysts.

However, there is always a chance that a transfer is taking place, even if it not strong enough to completely flatten out the regression lines. In other words, Fig. 9 shows positive correlations that would have been even greater for individual interviews.

Based on our data, we can compare the regression lines for group and individual interviews. Fig. 10 shows the regression lines for: (a) interview experience and (b) requirements experience and analyst effectiveness measured during individual interviews. The correlation coefficients ($r = -0.23$ and $r = 0.031$, respectively) are small to medium and not significant ($p\text{-value} = 0.114$ and $p\text{-value} = 0.835$, respectively).

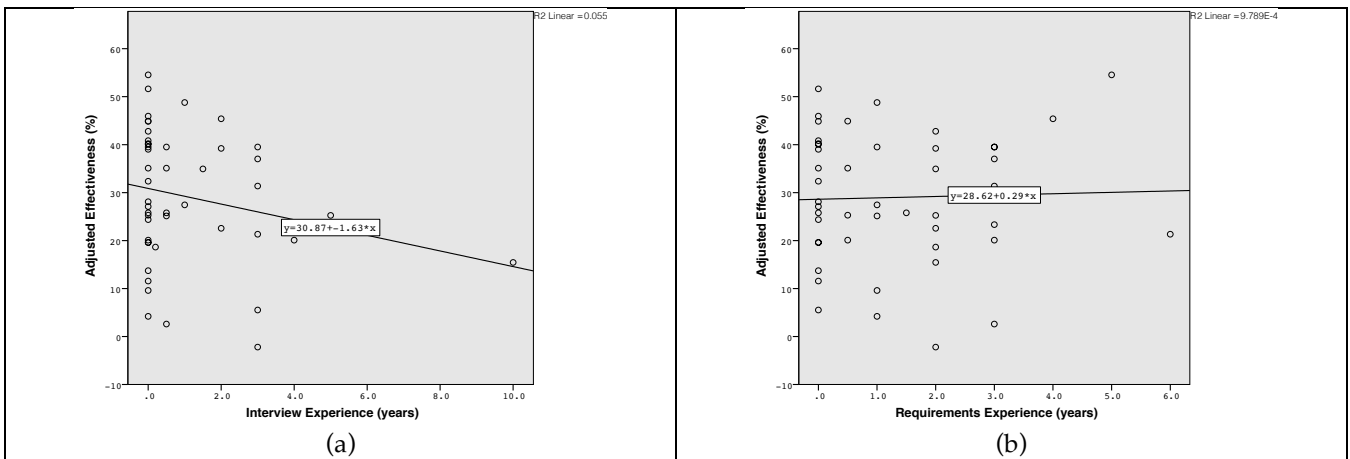

Fig. 10. Regression lines for individual interviews

In the case of requirements experience, it is clear that the associated correlation is greater in the group interview than in the individual interview. We believe that this is another argument against the transfer of information between experts and novices. The situation is more complex in the case of interview experience. Interview experience is greater for group than individual interviews (again, suggesting that there is no information transfer). The strange thing is that the correlation in this case is negative. We think that this might be due to a combination of factors such as experimental subjects overrating their interview experience and being under-motivated.

**G.4- DIFFERENCE BETWEEN HIGH AND LOW PERFORMERS**

TABLE 12
DIFFERENCES BETWEEN MOST AND LEAST EFFICIENT SUBJECTS – FP1

**Test Statistics**[a]

|  | Job Positions | Number of projects | Requirements Experience (months) | Professional Experience (years) | University Ranking | CGPA obtained in Bachelor courses |
|---|---|---|---|---|---|---|
| Mann-Whitney U | 22.500 | 40.000 | 37.500 | 36.000 | 40.000 | 13.500 |
| Wilcoxon W | 67.500 | 85.000 | 82.500 | 81.000 | 85.000 | 49.500 |
| Z | -1.657 | -.047 | -.265 | -.407 | -.044 | -1.680 |
| Asymp. Sig. (2-tailed) | .098 | .962 | .791 | .684 | .965 | .093 |
| Exact Sig. [2*(1-tailed Sig.)] | .113[b] | 1.000[b] | .796[b] | .730[b] | 1.000[b] | .094[b] |

a. Grouping Variable: GroupNum
b. Not corrected for ties.

TABLE 13
DIFFERENCES BETWEEN MOST AND LEAST EFFICIENT SUBJECTS) – UP1

**Test Statistics**[a]

|  | Job Positions | Number of projects | Requirements Experience (months) | Professional Experience (years) | University Ranking | CGPA obtained in Bachelor courses |
|---|---|---|---|---|---|---|
| Mann-Whitney U | 47.500 | 46.500 | 60.000 | 69.500 | 50.500 | 27.000 |
| Wilcoxon W | 138.500 | 137.500 | 138.000 | 160.500 | 141.500 | 55.000 |
| Z | -1.146 | -.820 | -.370 | -.473 | -1.502 | -1.042 |
| Asymp. Sig. (2-tailed) | .252 | .412 | .712 | .636 | .133 | .297 |
| Exact Sig. [2*(1-tailed Sig.)] | .284[b] | .431[b] | .740[b] | .650[b] | .137[b] | .328[b] |

a. Grouping Variable: GroupNum

# APPENDIX H: EFFECT OF EXPERIENCE DEPENDING ON DOMAIN ITEM TYPE

Below we report the regression models for each item type defining the problem domain (processes, concepts and requirements). We present both problem domains (UP1 and FP1) separately.

## H.1- UP1 PROBLEM DOMAIN

### TABLE 14
### PROCESSES

**Coefficients**

| Model | | Unstandardized Coefficients B | Unstandardized Coefficients Std. Error | Standardized Coefficients Beta | t | Sig. | Collinearity Statistics B | Collinearity Statistics Std. Error |
|---|---|---|---|---|---|---|---|---|
| 1 | (Constant) | 4.133 | .427 | | 9.686 | .000 | | |
| | Interview Experience (years) | .074 | .132 | .121 | .566 | .574 | .318 | 3.145 |
| | Requirements Experience (years) | .110 | .120 | .241 | .913 | .365 | .209 | 4.787 |
| | Development Experience (years) | .116 | .088 | .210 | 1.318 | .192 | .574 | 1.742 |
| | Professional Experience (years) | -.116 | .102 | -.294 | -1.144 | .257 | .221 | 4.535 |

a. Dependent variable: AdjustedProcesses (#)  $R^2 = .070$ Sig. = .314 N = 69 Expected-N= 82

### TABLE 15
### CONCEPTS

**Coefficients**

| Model | | Unstandardized Coefficients B | Unstandardized Coefficients Std. Error | Standardized Coefficients Beta | t | Sig. | Collinearity Statistics B | Collinearity Statistics Std. Error |
|---|---|---|---|---|---|---|---|---|
| 1 | (Constant) | 7.424 | .720 | | 10.310 | .000 | | |
| | Interview Experience (years) | .071 | .222 | .069 | .318 | .751 | .318 | 3.145 |
| | Requirements Experience (years) | .174 | .203 | .231 | .858 | .394 | .209 | 4.787 |
| | Development Experience (years) | .049 | .148 | .054 | .333 | .740 | .574 | 1.742 |
| | Professional Experience (years) | -.129 | .171 | -.197 | -.752 | .455 | .221 | 4.535 |

a. Dependent variable: Adjusted Concepts (#)  $R^2 = .177$ Sig. = .723 N = 69 Expected-N=82

### TABLE 16
### REQUIREMENTS

**Coefficients**

| Model | | Unstandardized Coefficients B | Unstandardized Coefficients Std. Error | Standardized Coefficients Beta | t | Sig. | Collinearity Statistics B | Collinearity Statistics Std. Error |
|---|---|---|---|---|---|---|---|---|
| 1 | (Constant) | 1.204 | .735 | | 1.637 | .107 | | |
| | Interview Experience (years) | .170 | .227 | .157 | .749 | .457 | .318 | 3.145 |
| | Requirements Experience (years) | -.024 | .207 | -.030 | -.114 | .910 | .209 | 4.787 |
| | Development Experience (years) | .042 | .152 | .043 | .277 | .783 | .574 | 1.742 |
| | Professional Experience (years) | -.295 | .175 | -.425 | -1.686 | .097 | .221 | 4.535 |

a. Dependent variable: Adjusted Requirements (#)  $R^2 = .104$ Sig. = .130 N = 69 Expected-N=82

## H.2- FP1 Problem Domain

### TABLE 17
#### Processes

**Coefficients**

| | | Unstandardized Coefficients | | Standardized Coefficients | t | Sig. | Collinearity Statistics | |
|---|---|---|---|---|---|---|---|---|
| Model | | B | Std. Error | Beta | | | B | Std. Error |
| 1 | (Constant) | 4.366 | 1.106 | | 3.947 | .001 | | |
| | Interview Experience (years) | 1.010 | .460 | .431 | 2.194 | .038 | .773 | 1.294 |
| | Requirements Experience (years) | .738 | .483 | .316 | 1.528 | .140 | .698 | 1.432 |
| | Development Experience (years) | -.123 | .389 | -.071 | -.316 | .755 | .597 | 1.674 |
| | Professional Experience (years) | -.554 | .263 | -.459 | -2.107 | .046 | .631 | 1.585 |

a. Dependent variable: AdjustedProcesses (#)    $R^2 = .283$ Sig. $= .082$ N $= 29$ Expected-N$= 82$

### TABLE 18
#### Concepts

**Coefficients**

| | | Unstandardized Coefficients | | Standardized Coefficients | t | Sig. | Collinearity Statistics | |
|---|---|---|---|---|---|---|---|---|
| Model | | B | Std. Error | Beta | | | B | Std. Error |
| 1 | (Constant) | 4.488 | .713 | | 6.296 | .000 | | |
| | Interview Experience (years) | .395 | .297 | .279 | 1.330 | .196 | .773 | 1.294 |
| | Requirements Experience (years) | .545 | .311 | .386 | 1.752 | .093 | .698 | 1.432 |
| | Development Experience (years) | -.110 | .251 | -.104 | -.437 | .666 | .597 | 1.674 |
| | Professional Experience (years) | -.130 | .169 | -.178 | -.768 | .450 | .631 | 1.585 |

a. Dependent variable: AdjustedConcepts (#)    $R^2 = .185$ Sig. $= .276$ N $= 29$ Expected-N$= 82$

### TABLE 19
#### Requirements

**Coefficients**

| | | Unstandardized Coefficients | | Standardized Coefficients | t | Sig. | Collinearity Statistics | |
|---|---|---|---|---|---|---|---|---|
| Model | | B | Std. Error | Beta | | | B | Std. Error |
| 1 | (Constant) | 6.223 | 1.441 | | 4.319 | .000 | | |
| | Interview Experience (years) | 1.145 | .600 | .384 | 1.909 | .068 | .773 | 1.294 |
| | Requirements Experience (years) | .486 | .629 | .164 | .772 | .448 | .698 | 1.432 |
| | Development Experience (years) | -.098 | .507 | -.044 | -.192 | .849 | .597 | 1.674 |
| | Professional Experience (years) | -.773 | .342 | -.504 | -2.260 | .033 | .631 | 1.585 |

a. Dependent variable: AdjustedRequirements (#)    $R^2 = .248$ Sig. $= .130$ N $= 29$ Expected-N$= 82$

# APPENDIX I: EFFECT OF EXPERIMENTS

Below we present the analysis of the MLR residuals for each problem domain and quasi-experiment.

## I.1- UP1 PROBLEM DOMAIN

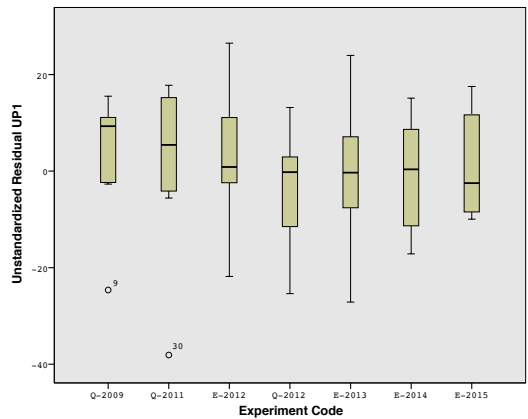

Fig. 11. Distribution of residuals of the eight quasi-experiments

TABLE 20

TEST OF BETWEEN-SUBJECTS EFFECTS

Dependent Variable: Non-standardized Residual for UP1

| Source | Type III Sum of Squares | df | Mean Square | F | Sig. |
|---|---|---|---|---|---|
| Corrected Model | 426.924a | 6 | 71.154 | .407 | .872 |
| Intercept | 23.333 | 1 | 23.333 | .133 | .716 |
| ExperimentCode | 426.924 | 6 | 71.154 | .407 | .872 |
| Error | 10842.020 | 62 | 174.871 | | |
| Total | 11268.944 | 69 | | | |
| Corrected Total | 11268.944 | 68 | | | |

a. R Squared = .038 (Adjusted R Squared = -.055)

TABLE 21

LEVENE'S TEST OF EQUALITY OF ERROR VARIANCES

Dependent Variable: Non-standardized Residual for UP1

| F | df1 | df2 | Sig. |
|---|---|---|---|
| .161 | 6 | 62 | .986 |

Tests the null hypothesis that the error variance of the dependent variable is equal across groups.
a. Design: Intercept + Experiment Code

## I.2- FP1 Problem Domain

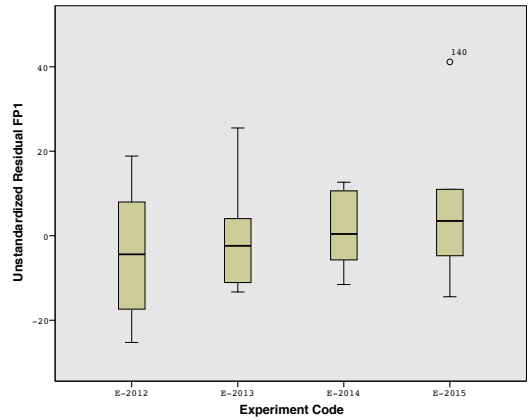

Fig. 12. Distribution of residuals of the four quasi-experiments

### TABLE 22
### Test of between-subjects effects

Dependent Variable: Non-standardized Residual for FP1

| Source | Type III Sum of Squares | df | Mean Square | F | Sig. |
|---|---|---|---|---|---|
| Corrected Model | 552.272a | 3 | 184.091 | .801 | .505 |
| Intercept | 25.628 | 1 | 25.628 | .112 | .741 |
| Experiment Code | 552.272 | 3 | 184.091 | .801 | .505 |
| Error | 5744.791 | 25 | 229.792 | | |
| Total | 6297.064 | 29 | | | |
| Corrected Total | 6297.064 | 28 | | | |

a. R Squared = .088 (Adjusted R Squared = -.022)

### TABLE 23
### Levene's test of equality of error variances

Dependent Variable: Non-standardized Residual for FP1

| F | df1 | df2 | Sig. |
|---|---|---|---|
| .477 | 3 | 25 | .701 |

Tests the null hypothesis that the error variance of the dependent variable is equal across groups.
a. Design: Intercept + Experiment Code